\documentclass[pra,aps,showpacs,twocolumn,twoside,superscriptaddress]{revtex4}
\usepackage{amsmath,amsfonts,amssymb,caption,color,epsfig,graphics,graphicx,latexsym,mathrsfs,revsymb,theorem,url,verbatim,epstopdf}

\usepackage{hyperref}

\newtheorem{definition}{Definition}
\newtheorem{proposition}[definition]{Proposition}
\newtheorem{lemma}[definition]{Lemma}

\newtheorem{theorem}[definition]{Theorem}
\newtheorem{corollary}[definition]{Corollary}
\newtheorem{conjecture}[definition]{Conjecture}

\newtheorem{remark}[definition]{Remark}
\newtheorem{example}[definition]{Example}
\newtheorem{question}[definition]{Question}

\def\squareforqed{\hbox{\rlap{$\sqcap$}$\sqcup$}}
\def\qed{\ifmmode\squareforqed\else{\unskip\nobreak\hfil
\penalty50\hskip1em\null\nobreak\hfil\squareforqed
\parfillskip=0pt\finalhyphendemerits=0\endgraf}\fi}
\def\endenv{\ifmmode\;\else{\unskip\nobreak\hfil
\penalty50\hskip1em\null\nobreak\hfil\;
\parfillskip=0pt\finalhyphendemerits=0\endgraf}\fi}
\newenvironment{proof}{\noindent \textbf{{Proof.~} }}{\qed}
\def\Dbar{\leavevmode\lower.6ex\hbox to 0pt
{\hskip-.23ex\accent"16\hss}D}
\makeatletter
\def\url@leostyle{%
  \@ifundefined{selectfont}{\def\UrlFont{\sf}}{\def\UrlFont{\small\ttfamily}}}
\makeatother
\def\bcj{\begin{conjecture}}
\def\ecj{\end{conjecture}}
\def\bcr{\begin{corollary}}
\def\ecr{\end{corollary}}
\def\bd{\begin{definition}}
\def\ed{\end{definition}}
\def\bea{\begin{eqnarray}}
\def\eea{\end{eqnarray}}
\def\bem{\begin{enumerate}}
\def\eem{\end{enumerate}}
\def\bex{\begin{example}}
\def\eex{\end{example}}
\def\bim{\begin{itemize}}
\def\eim{\end{itemize}}
\def\bl{\begin{lemma}}
\def\el{\end{lemma}}
\def\bpf{\begin{proof}}
\def\epf{\end{proof}}
\def\bpp{\begin{proposition}}
\def\epp{\end{proposition}}
\def\bqu{\begin{question}}
\def\equ{\end{question}}
\def\br{\begin{remark}}
\def\er{\end{remark}}
\def\bt{\begin{theorem}}
\def\et{\end{theorem}}

\def\btb{\begin{tabular}}
\def\etb{\end{tabular}}

\newcommand{\nc}{\newcommand}

\def\e{\epsilon}

\def\s{\sigma}

 \nc{\bA}{{\bf A}} \nc{\bB}{{\bf B}} \nc{\bC}{{\bf C}}
 \nc{\bD}{{\bf D}} \nc{\bE}{{\bf E}} \nc{\bF}{{\bf F}}
 \nc{\bG}{{\bf G}} \nc{\bH}{{\bf H}} \nc{\bI}{{\bf I}}
 \nc{\bJ}{{\bf J}} \nc{\bK}{{\bf K}} \nc{\bL}{{\bf L}}
 \nc{\bM}{{\bf M}} \nc{\bN}{{\bf N}} \nc{\bO}{{\bf O}}
 \nc{\bP}{{\bf P}} \nc{\bQ}{{\bf Q}} \nc{\bR}{{\bf R}}
 \nc{\bS}{{\bf S}} \nc{\bT}{{\bf T}} \nc{\bU}{{\bf U}}
 \nc{\bV}{{\bf V}} \nc{\bW}{{\bf W}} \nc{\bX}{{\bf X}}
 \nc{\bZ}{{\bf Z}}

\nc{\cA}{{\cal A}} \nc{\cB}{{\cal B}} \nc{\cC}{{\cal C}}
\nc{\cD}{{\cal D}} \nc{\cE}{{\cal E}} \nc{\cF}{{\cal F}}
\nc{\cG}{{\cal G}} \nc{\cH}{{\cal H}} \nc{\cI}{{\cal I}}
\nc{\cJ}{{\cal J}} \nc{\cK}{{\cal K}} \nc{\cL}{{\cal L}}
\nc{\cM}{{\cal M}} \nc{\cN}{{\cal N}} \nc{\cO}{{\cal O}}
\nc{\cP}{{\cal P}} \nc{\cQ}{{\cal Q}} \nc{\cR}{{\cal R}}
\nc{\cS}{{\cal S}} \nc{\cT}{{\cal T}} \nc{\cU}{{\cal U}}
\nc{\cV}{{\cal V}} \nc{\cW}{{\cal W}} \nc{\cX}{{\cal X}}
\nc{\cZ}{{\cal Z}}

\nc{\hA}{{\hat{A}}} \nc{\hB}{{\hat{B}}} \nc{\hC}{{\hat{C}}}
\nc{\hD}{{\hat{D}}} \nc{\hE}{{\hat{E}}} \nc{\hF}{{\hat{F}}}
\nc{\hG}{{\hat{G}}} \nc{\hH}{{\hat{H}}} \nc{\hI}{{\hat{I}}}
\nc{\hJ}{{\hat{J}}} \nc{\hK}{{\hat{K}}} \nc{\hL}{{\hat{L}}}
\nc{\hM}{{\hat{M}}} \nc{\hN}{{\hat{N}}} \nc{\hO}{{\hat{O}}}
\nc{\hP}{{\hat{P}}} \nc{\hR}{{\hat{R}}} \nc{\hS}{{\hat{S}}}
\nc{\hT}{{\hat{T}}} \nc{\hU}{{\hat{U}}} \nc{\hV}{{\hat{V}}}
\nc{\hW}{{\hat{W}}} \nc{\hX}{{\hat{X}}} \nc{\hZ}{{\hat{Z}}}

\nc{\hn}{{\hat{n}}}

\def\diag{\mathop{\rm diag}}
\def\dim{\mathop{\rm Dim}}

\def\max{\mathop{\rm max}}
\def\min{\mathop{\rm min}}

\def\rank{\mathop{\rm rank}}

\def\bigox{\bigotimes}
\def\dg{\dagger}

\def\op{\oplus}
\def\ox{\otimes}

\def\ra{\rightarrow}

\def\sue{\subseteq}

\newcommand{\bra}[1]{\langle#1|}
\newcommand{\ket}[1]{|#1\rangle}
\newcommand{\proj}[1]{| #1\rangle\!\langle #1 |}
\newcommand{\ketbra}[2]{|#1\rangle\!\langle#2|}

\def\Dbar{\leavevmode\lower.6ex\hbox to 0pt
{\hskip-.23ex\accent"16\hss}D}
\begin{document}
\title{Decomposition of bipartite and multipartite unitary gates into the product of controlled unitary gates}

\author{Lin Chen}
\affiliation{Department of Mathematics, Beijing University of Aeronautics and Astronautics, Beijing 100191, P. R. China}
\affiliation{Singapore University of Technology and Design, 20 Dover Drive, Singapore 138682}
\author{Li Yu}\email{liyu@nii.ac.jp}
\affiliation{Singapore University of Technology and Design, 20 Dover Drive, Singapore 138682}
\affiliation{National Institute of Informatics, 2-1-2 Hitotsubashi, Chiyoda-ku, Tokyo 101-8430, Japan}

\date{\today}


\begin{abstract}
We show that any unitary operator on the $d_A\times d_B$ system ($d_A\ge 2$) can be decomposed into the product of at most $4d_A-5$ controlled unitary operators. The number can be reduced to $2d_A-1$ when $d_A$ is a power of two. We also prove that three controlled unitaries can implement a bipartite complex permutation operator, and discuss the connection to an analogous result on classical reversible circuits. We further show that any $n$-partite unitary on the space $\mathbb{C}^{d_1}\otimes\cdots\otimes\mathbb{C}^{d_n}$ is the product of at most $[2\prod^{n-1}_{j=1}(2d_j-2)-1]$ controlled unitary gates, each of which is controlled from $n-1$ systems. We also decompose any bipartite unitary into the product of a simple type of bipartite gates and some local unitaries. We derive dimension-independent upper bounds for the CNOT-gate cost or entanglement cost of bipartite permutation unitaries (with the help of ancillas of fixed size) as functions of the Schmidt rank of the unitary. It is shown that such costs under a simple protocol are related to the log-rank conjecture in communication complexity theory via the link of nonnegative rank.

\end{abstract}

\maketitle

%\tableofcontents

\section{introduction}\label{sec:intro}

The implementation of unitary operations is a key task in quantum information processing. Unitary operators can be implemented by passive linear optical devices \cite{Reck94}. It is known that any unitary operation on two or more parties can be decomposed into the product of controlled unitary gates \cite{bry02,blb05}.  Two-qubit controlled unitaries can be implemented with high coherence and dynamical coupling \cite{cnr14}. Suppose that a bipartite unitary $U$ on systems $A$,$B$ is the product of $k$ bipartite controlled unitaries, interspersed with local unitaries \cite{note1}. We call the integer $k$ as the \emph{bipartite depth} of the circuit under the bipartite cut $A$-$B$. The depth, width and total number of basic gates are often quantities of interest in quantum circuit design, where the basic gates refer to some fixed type of two-qubit gates such as the controlled-NOT (CNOT) gate. For implementing the same unitary operation, it is conceivable that there may be a tradeoff between the depth and the total number of basic gates. Nonetheless the bipartite depth does give an upper bound for the total number of basic gates, as discussed in Sec.~\ref{sec:elementary} of this paper.  The nonlocal gates need much longer time than local gates to implement, because the systems may be far from each other. Then the bipartite depth is a rough measure of time needed by the circuit. By allowing local unitary freedom in the definition of controlled unitaries (in Sec.~\ref{sec:pre}), from now on we will drop the phrase ``interspersed with local unitaries'' from the definition of the bipartite depth.

We define the \emph{bipartite depth of a given bipartite unitary $U$} as the minimum bipartite depth among all unitary circuits for $U$ that do not use ancillas. Formally, it is
\begin{eqnarray}
\label{eq:cu}
c(U) := \min\{k | U=U_1U_2\cdots U_k,~~U_i\in \cS\},
\end{eqnarray}
where $\cS$ is the set of bipartite controlled unitaries on the same space that $U$ acts on. Studying the bounds for $c(U)$ and the corresponding decomposition of $U$ is the main problem in this paper. Indeed, it is a special case of the problem of quantum circuit decomposition using \emph{general} controlled unitaries with the help of local unitaries. It is special in the sense that there are only two systems but the general problem allows many systems. There has been study on decompositions using CNOT or other two-qubit controlled gates, or specific classes of two-qudit controlled gates \cite{bry02,blb05,bbo06,dw13}. For example, Shende \textit{et al.} \cite{Shende06} shows that any three-qubit unitary can be written as the product of 20 CNOT gates and some one-qubit unitaries. Another motivation to study the problem is to better understand the structure of nonlocal unitaries and the resources needed to implement them, see the comment just before Section~\ref{ssec1}.

We restrict to bipartite controlled gates as the type of nonlocal gates in the definition of bipartite depth for the following reasons. First, it is easy to define, and a smaller class of gates seems not powerful enough. It is hard to find a larger class of easily definable gates that do not include all bipartite gates. The Fourier hierarchy \cite{Shi2005335} concerns the number of tensor products of Hadamard gates in a circuit that also contains basis-preserving gates. The basis-preserving gates are also called the complex permutation gates, and are discussed later in this paper. They permute among computational-basis states and apply a phase to each state. However the basis-preserving gates are generally nonlocal with respect to a bipartite partition of the qubits. If we modify the definition of Fourier hierarchy and apply it to the bipartite scenario so as to allow some finite set of bipartite gates and arbitrary local gates, then such a set of bipartite gates would have a discrete set of entangling power, which is not desirable for defining a smooth depth measure. Second, the controlled unitaries are analogous to some components in protocols with local operations and classical communication (LOCC). They are a major type of protocols studied in quantum information theory. The LOCC protocols often allow projective measurements on some subsystems. A projective measurement and the subsequent classically controlled unitary operations can be made part of a coherent quantum circuit by rewriting them as a controlled unitary. Thus our measure is analogous to the rounds of classical communications in such protocols.

Generally we consider unitaries acting on $d_A\times d_B$ dimensional systems. The results of \cite{bry02,blb05} imply that $c(U)\le \mu d_A^4$ when $d_A=d_B$, where $\mu$ is a positive constant, and the type of bipartite controlled gates used are limited to controlled-increment gates. In Theorem \ref{thm:u_controlled}, we obtain a tighter bound $c(U)\le 4d_A-5$ for arbitrary $d_A,d_B$ at the cost of allowing the use of arbitrary controlled-unitary gates in the decomposition. The same theorem shows that the bound can be further reduced to $2d_A-1$ when $d_A$ is a power of $2$.  We also prove that $c(U)\le 3$ when $U$ is a complex permutation matrix in Theorem~\ref{thm:u_controlled2}, based on the concept of absolute singularity studied in Lemma \ref{le:absolutesingular}. This result is applied to classical reversible circuits \cite{Shende03,sm14} in Corollary~\ref{cr:u_classical}. The above results are based on the sandwich form of bipartite unitaries, constructed in Definition \ref{df:sandwich} and Lemma \ref{le:2xddecomp}.
We further generalize our observation to multipartite systems based on the generalized sandwich form. We show that any $n$-partite unitary on the space $\mathbb{C}^{d_1}\ox\cdots\ox\mathbb{C}^{d_n}$ has a generalized $[2\prod^{n-1}_{j=1}(2d_j-2)-1]$-sandwich form in Proposition \ref{pp:multi_controlled}. We also propose a more efficient generalized sandwich form for $n=4$ in Proposition \ref{pp:multi_controlled2}. In Proposition \ref{pp:multi_controlled3}, we show that any $n$-partite complex permutation unitary has a generalized $(2^n-1)$-sandwich form composed of controlled-complex-permutation unitaries.

We also discuss the decomposition of any unitary gate using ``standard'' gates proposed in Definition \ref{def:ele1}. They effectively only act on two qubits as controlled unitaries, and may be more easily carried out in experiments. We show that any bipartite unitary is the product of $2(d_A-1)^2 \lfloor \frac{d_B}{2}\rfloor
+(2d_A-3)(d_B-1) \lfloor \frac{d_A}{2}\rfloor$ standard gates interspersed with local unitaries in Proposition \ref{pp:elem}. The number reduces to three for $d_A=d_B=2$, which is the smallest number of controlled unitaries needed for the decomposition of two-qubit unitary gates \cite{vw04}.
In Sec.~\ref{sec:Schmidt} we discuss the relationship between the Schmidt rank of the unitary and the number of controlled unitaries needed to decompose it. We give a class of examples where the number of controlled unitaries is upper bounded by a constant, but the Schmidt rank of the target unitary is arbitrarily large.

The rest of the paper is organized as follows. In Sec.~\ref{sec:pre} we introduce some definitions and preliminary knowledge. In Sec.~\ref{sec:bipartite} we study the decomposition of bipartite unitary operators using controlled unitaries, and comment on the connections with results in the literature. In Sec.~\ref{sec:multi} we define the ``controlled-type'' multipartite unitaries and discuss the decomposition of multipartite operators into the product of these gates. We also show that three controlled-permutation matrices are enough to decompose any complex permutation matrix. In Sec.~\ref{sec:elementary} we define the standard gates  and discuss the decomposition of bipartite unitaries using these gates and local unitaries. In Sec.~\ref{sec:Schmidt} we discuss the relationship between the Schmidt rank of the unitary and the form of the decomposition, and we discuss bipartite permutation unitaries in particular. In Sec.~\ref{sec:ancilla} we discuss the use of local ancillas. We conclude in Sec.~\ref{sec:con}.

\section{Preliminaries}\label{sec:pre}

In this section we introduce the preliminary knowledge used in the paper. Denote the computational-basis states of the bipartite Hilbert space $\cH=\cH_A\ox\cH_B$ by $\ket{i,j},i=1,\cdots,d_A$, $j=1,\cdots,d_B$. Let
$I_A$ and $I_B$ be the identity
operators on the spaces $\cH_A$ and $\cH_B$, respectively. Any bipartite unitary gate $U$ acting on $\cH$ has \emph{Schmidt rank} (denoted as ${\rm Sch}(U)$) equal to $n$ if there is an expansion of the form $U=\sum^n_{j=1}A_j \ox B_j$ where the $d_A\times d_A$ matrices $A_1,\cdots,A_n$ are linearly independent, and the $d_B\times d_B$ matrices $B_1,\cdots,B_n$ are also linearly independent. An equivalent definition is in \cite{Nielsen03,Tyson03}, where it is called the operator-Schmidt rank.
Next, $U$ is a \textit{controlled unitary gate}, if $U$ is
 equivalent to $\sum^{d_A}_{j=1}\proj{j}\ox U_j$ or
$\sum^{d_B}_{j=1}V_j \ox \proj{j}$ via local unitaries. To be specific, $U$ is a controlled unitary from the $A$ or $B$ side, respectively.
In particular, $U$ is controlled in the computational basis from the $A$ side if $U=\sum^{d_A}_{j=1}\proj{j}\ox U_j$. Bipartite unitary gates of Schmidt rank two or three are in fact controlled unitaries \cite{cy13,cy14,cy14ap}.
We have generalized controlled unitaries to block-controlled unitary gates \cite{cy14}. We split the space $\cH_A$ into a direct sum:
$\cH_A=\op^m_{i=1} \cH_i$, $m>1$, $\dim\cH_i=m_i$, and
$\cH_i\perp\cH_j$ for distinct $i,j=1,\cdots,m$. Then $U$ is
a \textit{block-controlled unitary (BCU) gate controlled from the A
side}, if $U$ is locally equivalent to $\sum^m_{i=1}
\sum^{m_i}_{j,k=1} \ketbra{u_{ij}}{u_{ik}}\ox V_{ijk}$ where
$\{\ket{u_{i,1}},\cdots,\ket{u_{i,m_i}}\}$ is an orthonormal basis of $\cH_i$. Note that the $V_{ijk}$ are not necessarily unitary. By definition every controlled unitary with $d_A,d_B\ge2$ is a BCU. The BCU will be used in the proof of Theorem \ref{thm:u_controlled}, as well as in the decomposition of any bipartite unitary into the product of three BCUs in Corollary \ref{cr:bcu}.

\section{Decomposition of bipartite unitary operators}

\label{sec:bipartite}

It is known \cite{vw04} that three controlled gates are sufficient and necessary for the decomposition of a general two-qubit unitary, and there is always a decomposition using $3$ CNOT gates and some one-qubit unitaries. For implementing a two-qubit SWAP gate by local unitaries and some number of CNOT gates without the use of ancillas (this condition of no ancillas is implied throughout the paper unless stated otherwise), three CNOT gates are necessary and sufficient \cite{vw04}. We generalize this fact to the SWAP gates of arbitrary dimension.

\bl
\label{le:swap}
Denote the two-qudit SWAP gate acting on $d\times d$ system as SWAP$_d$. Then
\\
(i)
the product of the SWAP$_d$ gate and any controlled unitary has Schmidt rank $d^2$;
\\
(ii)
For implementing a SWAP$_d$ gate by local unitaries and some number of controlled unitary gates, three controlled unitaries are necessary and sufficient.
\el
\bpf
(i) There are orthonormal bases of $\cH_A$ and $\cH_B$ (denoted by $\{\ket{i}\}_A$ and $\{\ket{j}\}_B$) such that the matrix representation of the SWAP$_d$ gate in such bases has elements of the form $\bra{i}_A\bra{j}_B U\ket{k}_A \ket{l}_B=\delta_{il}\delta_{jk}$. Because the SWAP$_d$ gate effectively performs the physical swap of two systems, which is basis-independent, the above particular matrix representation is invariant under simultaneous unitary similarity transform (simultaneous unitary change of basis) on the two local systems. Then assertion (i) follows from straightforward computation, by writing the matrix for the SWAP$_d$ gate in the form above and assuming one of the local bases is the local controlling basis for the controlled unitary.

(ii) Any controlled unitary on $\cH$ has Schmidt rank at most $d$. It follows from assertion (i) that the SWAP$_d$ gate is the product of at least three controlled unitaries. It is known that the SWAP$_d$ gate is the product of three controlled unitary gates \cite{gc13}. So assertion (ii) holds.
This completes the proof.
\epf

\smallskip
For the two-qubit SWAP gate, using the general controlled unitaries in its decomposition does not save any controlled unitary compared to using CNOT gates. One might expect that this is the general case, i.e.,  the implementation of a bipartite unitary is the same when we use controlled unitaries or only CNOT gates. However, the two-qubit gate $\exp(ia\s_1\ox \s_1)$ with the Pauli matrix $\s_1=\left(
                   \begin{array}{cc}
                     0 & 1 \\
                     1 & 0 \\
                     \end{array}
                 \right)$ any $a\ne k\pi/4$, $k\in \mathbb{Z}$ cannot be implemented using one CNOT gate and single qubit gates only, since the entangling power of such gate is not equal to that of the CNOT gate.  We will show in Theorem \ref{thm:u_controlled} that for the general $d\times d$ bipartite system that using controlled unitaries might be better than the $d$-dimensional CNOT gates, in the sense that they require fewer such two-qudit gates.
For this purpose we introduce a special decomposition of bipartite unitaries.
\bd
\label{df:sandwich}
(i) We refer to the \textit{m-sandwich form} of a bipartite unitary $U$, in the sense that $U=U_1U_2\cdots U_m$, where each $U_i$ is a controlled unitary, being controlled in the computational basis on the respective Hilbert space, and the party that does the controlling alternates between A for odd $i$ and B for even $i$.
\\
(ii) We refer to the \textit{m-A form} of a bipartite unitary $U$, in the sense that $U=U_1U_2\cdots U_m$, where any $U_i$ is a controlled unitary controlled from the $A$ side.
\ed
Using this definition we present the following result as the first step to our question.

\bl
\label{le:2xddecomp}
(i) Any $2\times d_B$ unitary has a 3-sandwich form;
\\
(ii) Any $2\times d_B$ unitary has a 3-A form;
\\
(iii) There exists a $2\times 2$ unitary that cannot be the product of two controlled unitaries.
\el
\bpf
(i) For any $2\times d_B$ unitary $M$, there are two local unitaries $E,F$ on $\cH_B$ such that $M=(I_A\ox E)U(I_A\ox F)$,
where $U=\sum^1_{i,j=0}\ketbra{i}{j}\ox U_{ij}$ and $U_{00}$ is a $d_B\times d_B$ diagonal matrix. Since $U$ is unitary, the columns of $U_{10}$ are pairwise orthogonal, and the rows of $U_{01}$ are also pairwise orthogonal. Let $V,W$ be two $d_B\times d_B$ unitaries such that both $VU_{10}$ and $U_{01}W$ are diagonal matrices with all elements real and non-negative. Let $U_1=\proj{0}\ox I_B + \proj{1}\ox V$ and $U_2=\proj{0}\ox I_B + \proj{1}\ox W$ be two controlled unitaries from the $A$ side, we have
\bea
U_3 =U_1 U U_2 =
\left(
                   \begin{array}{cc}
                     U_{00} & U_{01}W \\
                     VU_{10} & VU_{11}W
                   \end{array}
                 \right).
\eea
Since $U$ is unitary, we have $U_{01}W=VU_{10}$. The matrix $U_3$ is a $2\times d_B$ bipartite unitary of Schmidt rank at most 3, so it is a controlled unitary from the $B$ side \cite{cy13,cy14}. We have proved that $U$ is the product of three controlled unitaries $U_1^\dg, U_3,$ and $U_2^\dg$. There exist suitable local unitaries $S=I_A\ox X_B$ and $T=I_A\ox Y_B$, so that $S U_3 T$ is controlled in the computational basis of $\cH_B$. Hence $U=(U_1^\dg S^\dg) (S U_3 T) (T^\dg U_2^\dg)$ is a decomposition with each of the three parts controlled in the computational basis of $\cH_A$ or $\cH_B$.  Therefore $M=\left[(I_A\ox E) (U_1^\dg S^\dg)\right] (S U_3 T) \left[(T^\dg U_2^\dg)  (I_A\ox F)\right]$ is exactly a 3-sandwich form.  Hence the assertion holds.

(ii) From the proof of (i), we know that any $2\times d_B$ unitary $U$ has a 3-sandwich form. Let $U=V_1V_2V_3$ where $V_1,V_3$ are controlled unitaries controlled in the computational basis of $\cH_A$, and $V_2$ is a controlled unitary controlled in the computational basis of $\cH_B$. Since $V_2$ is controlled in the computational basis of $\cH_B$, one can write $V_2=\sum^1_{i,j=0} \ketbra{i}{j}\ox V_{ij}$ where all $V_{ij}$ are diagonal matrices. By multiplying $V_2$ with two suitable diagonal controlled unitaries respectively from the left and right side, we can make all entries of $V_{00},V_{01}$ and $V_{10}$ real and non-negative, and the entries of $V_{11}$ real and non-positive. Since $V_2$ is unitary, we have $V_{00}=-V_{11}$ and $V_{01}=V_{10}$. So $V_2$ has Schmidt rank at most two. It is controlled from the $A$ side \cite{cy13}. The inverse of all diagonal unitary operators taken above are also diagonal, so they can be absorbed by $V_1$ and $V_3$. The latter are still controlled unitaries from the $A$ side in the computational basis. So $U=V_1V_2V_3$ is a 3-A form and the assertion holds.

(iii) The assertion follows from Lemma \ref{le:swap},  which shows that the two-qubit SWAP gate is a product of three controlled unitaries, and no fewer.
This completes the proof.
\epf

\smallskip
When $d_B=2$ namely the unitary acts on two-qubit states, assertion (ii) has been proved as the statement that any two-qubit unitary has the so-called canonical form \cite{kbg01,kc01}. It has been shown that any two-qubit unitary does not have Schmidt rank three \cite{Nielsen03}. For readers' reference, the Schmidt-rank-three multiqubit unitary has been investigated and constructed in \cite{cy13,cy14ap}.

Now we are in a position to give an upper bound of $c(U)$ and the associated method of decomposing the bipartite unitary $U$.
\bt\label{thm:u_controlled}
Let $U$ be a bipartite unitary on the $d_A\times d_B$ system. Then
\\
(i) $U$ has a $(2^{\lceil \log_2 d_A\rceil+1}-1)$-sandwich form. Hence
\bea
\label{eq:upperbound}
c(U) \le 2^{\lceil \log_2 d_A\rceil+1}-1 \le 4d_A-5,
\eea
for any $d_A\ge2$. In particular, $c(U)\le 2d_A-1$ when $d_A$ is an integer power of $2$.
\\
(ii) If all bipartite unitaries on the $d_A\times d_B$ system with odd $d_A\ge3$ have $(2d_A-1)$-sandwich forms, then $U$ has a $(2d_A-1)$-sandwich form for any even $d_A\ge2$.
\et
\bpf
(i) One can easily show that the second inequality in \eqref{eq:upperbound} holds. In particular its equality holds when $d_A=2^n+1$ with any nonnegative integer $n$. Since the first inequality in \eqref{eq:upperbound} and the last assertion of (i) both follow from the first assertion of (i), it is sufficient to prove the latter. The assertion is trivial if $d_A$ or $d_B=1$, so we assume $d_A,d_B\ge2$. The proof is by induction over $d_A$. The assertion for $d_A=2$ with any $d_B\ge 2$ is proven in Lemma~\ref{le:2xddecomp}. In the following we prove the assertion for a fixed $d_A\ge 3$, under the induction hypothesis that the $k\times d_B$ bipartite unitary with any $2\le k\le d_A-1$ and $d_B\ge 2$ has a $g(k)$--sandwich form, where for any positive integer $j$ we define
\bea
\label{eq:gk}
g(j)=2^{\lceil \log_2 j\rceil+1}-1.
\eea
Let $\cH_{A_1},\cH_{A_2}\sue\cH_A$ be two subspaces spanned by the first $y~(y\le \lfloor d_A/2 \rfloor)$ and $2y$ computational basis kets, respectively. Let
$V=I_{A_1}\ox I_B + V'$ be a BCU where $V'$ is a bipartite unitary on the subspace $H=\cH_{A_1}^\perp\ox\cH_B$. Let
\bea
\label{eq:w=w'+poxi}
W=W' + I_{{A_2}^\perp}\ox I_B
\eea
be another BCU, where $W'$ is a bipartite unitary on the subspace $\cH_{A_2}\ox\cH_B$, and $I_{{A_2}^\perp}$ is the identity operator on the subspace $\cH_{A_2}^\perp$. We can find a suitable $V$, such that in the top $yd_B$ rows of the matrix product $UV$, the nonzero entries occur only in the first $2yd_B$ columns. Then we can find a suitable $W$ such that the matrix product
\bea
\label{eq:x}
X:=UVW=I_{A_1}\ox I_B + X',
\eea
where $X'$ is a unitary acting on $H$. So $X$ is a BCU controlled from the $A$ side, and
\bea
\label{eq:u=xwdgvdg}
U=XW^\dg V^\dg.
\eea
By regarding $W'$ as a  $2\times yd_B$ bipartite unitary and using Lemma \ref{le:2xddecomp}, we obtain that $W'$ has a 3-sandwich form. Let
\bea
\label{eq:w'dag}
(W')^\dg=CTD,
\eea
where $C,D$ are both the direct sum of two unitaries each of order $yd_B$, and
\bea
\label{eq:t}
T=\sum^{yd_B}_{i=1} W_i \ox \proj{i}
\eea
with some unitaries $W_i$ of order two. So $C,D$ and $T$ can all be regarded as $2y\times d_B$ bipartite unitaries on the subspace $\cH_{A_2}\ox\cH_B$. Using \eqref{eq:w=w'+poxi}, \eqref{eq:u=xwdgvdg}, and \eqref{eq:w'dag}, we have
\bea
\label{eq:xctdv}
U=X(CTD+I_{{A_2}^\perp}\ox I_B)V^\dg=(X \tilde C)\tilde T(\tilde D V^\dg),
\eea
where
\bea
\label{eq:tildec}
\tilde C=C+I_{{A_2}^\perp}\ox I_B,
\\
\label{eq:tilded}
\tilde D=D+I_{{A_2}^\perp}\ox I_B,
\\
\label{eq:tildet}
\tilde T=T+I_{{A_2}^\perp}\ox I_B.
\eea
It follows from \eqref{eq:t} that $T$ can be regarded as a controlled unitary on $\cH_{A_2}\ox\cH_B$, controlled from the $B$ side in the computational basis. This fact and \eqref{eq:tildet} imply that $\tilde{T}$ is a controlled unitary from the $B$ side in the computational basis. Next,
it follows from \eqref{eq:x} and \eqref{eq:tildec} that $X\tilde C$ is a BCU, i.e.,
\bea
\label{eq:xtildec}
X\tilde C=X_1+X_2,
\eea
where the bipartite unitaries $X_1$ and $X_2$ act on the subspaces $H^\perp$ and $H$, respectively. Since $\dim \cH_{A_1}=y$ and $\dim \cH_{A_1}^\perp=d_A-y$, they are both smaller than $d_A$ for any $y=1,2,\cdots,\lfloor d_A/2 \rfloor$. It follows from the induction hypothesis that $X_1$ and $X_2$ have $g(y)$ and $g(d_A-y)$-sandwich forms, respectively. We have two decomposition
\bea
\label{eq:x1}
X_1=\prod^{g(y)}_{i=1} X_{1,i},
~~~~
X_2=\prod^{g(d_A-y)}_{i=1} X_{2,i},
\eea
where for any odd and even $i$, the $X_{j,i}$ is a controlled unitary from the $A$ and $B$ side, respectively. Then so is $X_{1,i}+X_{2,i}$, because $X_{1,i}$ and $X_{2,i}$ act on the subspaces $H^\perp$ and $H$, respectively. It follows from \eqref{eq:gk} and the condition $y\le \lfloor d_A/2 \rfloor$ that $g(y)\le g(d_A-y)$. This inequality, \eqref{eq:xtildec} and \eqref{eq:x1} imply $X\tilde C=\prod^{g(y)}_{i=1} (X_{1,i}+X_{2,i})\cdot\prod^{g(d_A-y)}_{j=g(y)+1} (I_{A_1}\ox I_B+X_{2,j})$. These facts imply that $X\tilde C$ has a $g(d_A-y)$-sandwich form. Next using the same argument except that \eqref{eq:tildec} is replaced by \eqref{eq:tilded}, one can show that $\tilde D V^\dg$ also has a $g(d_A-y)$-sandwich form. Third it follows from \eqref{eq:gk} that $g(j)$ is odd for any positive integer $j$. Fourth in the paragraph below \eqref{eq:tildet}, we have shown that $\tilde T$ is a controlled unitary from the $B$ side in the computational basis. Applying these four facts to \eqref{eq:xctdv} implies that the unitary $U$ has an $x$-sandwich form where
\bea
\label{eq:x2}
x
&&=\min_{1\le y\le\lfloor d_A/2 \rfloor} \big( 2 g(d_A-y) + 1 \big)
\notag\\
&&= 2 g(\lceil d_A/2 \rceil) + 1
= g(d_A).
\eea
The last two equalities in \eqref{eq:x2} follow from \eqref{eq:gk}, and the fact that $\lceil \log_2 d_A \rceil = \lceil \log_2 (d_A + 1) \rceil$ for odd $d_A\ge3$. So \eqref{eq:x2} is exactly the first assertion of (i).

(ii) The proof is by induction over even $d_A\ge2$. The assertion for $d_A=2$ with any $d_B\ge 2$ is proven in Lemma~\ref{le:2xddecomp}. In the following we prove the assertion for a fixed even $d_A\ge 4$, under the induction hypothesis that the $k\times d_B$ bipartite unitary with any even $k\in[2, d_A-1]$ and $d_B\ge 2$ has a $(2k-1)$--sandwich form. One can verify that the argument from the paragraph below \eqref{eq:gk} to the second sentence below \eqref{eq:xtildec} still applies here. We choose $y=d_A/2$ in the argument. If $y$ is odd (respectively, even), then the condition in (ii) (respectively, the induction hypothesis) implies that $X_1$ and $X_2$ in \eqref{eq:xtildec} both have $(d_A-1)$-sandwich forms, respectively. Hence \eqref{eq:x1} and the subsequent paragraph hold, except that $g(j)$ is replaced by $2j-1$ for any positive integer $j$.
Since $d_A\ge2$ is even, applying these facts to \eqref{eq:xctdv} implies that the unitary $U$ has an $x$-sandwich form where
\bea
\label{eq:x=2d-1}
x
=2(d_A-1)+1=2d_A-1.
\eea
This completes the proof of assertion (ii).
\epf

\smallskip
We do not know whether the condition in Theorem \ref{thm:u_controlled} (ii) can be satisfied, and we leave it as an open problem. As a byproduct of the theorem, it follows from \eqref{eq:u=xwdgvdg} that
\bcr
\label{cr:bcu}
Any bipartite unitary is the product of three BCUs controlled from the $A$, $B$ and $A$ sides, respectively.
\ecr
It is known that any two-qubit BCU is a controlled unitary. Hence Lemma \ref{le:2xddecomp} (iii) implies that the two-qubit CNOT gate cannot be the product of only two BCUs. In other word, the upper bound three in Corollary \ref{cr:bcu} is tight.

The upper bound obtained in Theorem \ref{thm:u_controlled} is $4d_A-5$ and it is polynomially smaller than $4d_A^4$ obtained in \cite{bry02}. Compared to the latter, the implementation of a bipartite unitary by arbitrary controlled unitaries can indeed save quantum resources.   Since the systems $A$ and $B$ are symmetric in the problem, $4d_B-5$ is also an upper bound for the number of controlled gates.  We consider the optimality of the bound $4d_A-5$ under the assumptions that $d_A\le d_B$ and that the number of controlled gates is a function of $d_A$ only. By parameter counting, the $4d_A-5$ is already optimal up to a constant factor, because the entire unitary has $d_A^2 d_B^2$ free real parameters in it, and each controlled unitary from the $A$ side and controlled in the computational basis of $\cH_A$ has $d_A d_B^2$ free real parameters in it, while each controlled gate from the $B$ side has $d_B d_A^2$ free real parameters in it, less than what is in a controlled gate from the $A$ side (so that a larger number of these would be used if they are used instead of controlled gates from the $A$ side).  Note that for two adjacent controlled gates, we have overestimated the number of free parameters, since when they are both controlled from the $A$ side, the change of controlling basis on $\cH_A$ could be viewed as a change in either of the controlled gates, and generally, a bipartite diagonal gate between two adjacent controlled gates can be absorbed into any of the two adjacent controlled gates.  But such issues only affect the count above by a lower order factor.

We comment on the connection with the results in the literature. Our Lemma~\ref{le:2xddecomp}(i) in the special case that $d_B$ is an integer power of $2$ is the same as Theorem 10 of Shende \textit{et al.} \cite{Shende06} (see also \cite{pw94}).  Our Theorem~\ref{thm:u_controlled}(i) in the case that $d_A$ is an integer power of $2$ can also be derived by recursively applying Theorem 10 of \cite{Shende06} (the first step of recursion is illustrated in Theorem 11 of \cite{Shende06}, and note that a gate controlled by multiple qubits belonging to the same party is a controlled gate in our language).  We abbreviate the details here.  Therefore our result can be viewed as a generalization of the results in \cite{Shende06} to the general dimensions.  Based on our result, it may be possible to decompose any qudit circuit (with dimensions of qudits not required to be all equal) using controlled two-qudit unitaries.  The following Sec.~\ref{sec:multi} can be viewed as a step in this direction, but we do not decompose the gates fully there, allowing some gate controlled by multiple qudits. There may be some extensions of the techniques in \cite{Shende06} to the case of higher dimensional qudits that can help decompose such multiply-controlled gate. There are some papers on decomposition of qudit circuits, such as \cite{blb05,bbo06,dw13}. It is possible that the methods in those papers may be combined with the results in this paper to give a better upper bound of the number of two-qudit (controlled) gates needed. Apart from the application to circuit decomposition, the other potential application is to help study the nonlocal resource usage in implementing nonlocal unitaries. Here the usage of nonlocal resources is to be optimized, and the local resources such as local unitaries are deemed as cheap. Section~\ref{sec:elementary} is a step in this direction, but it only discusses the cost in terms of a particular type of nonlocal gate (whose implementation cost is upper bounded by a constant), and not in terms of the more conventional resources such as entanglement.

\subsection{Decomposition of complex permutation matrices}\label{ssec1}

The upper bound in Theorem \ref{thm:u_controlled} works for arbitrary bipartite unitaries, and it increases linearly with the dimension. One may expect to have a constant upper bound for some special bipartite unitaries. In this subsection we give such a bound for any \emph{complex permutation matrix} in Theorem \ref{thm:u_controlled2}. It is a unitary matrix with one and only one nonzero element on each row and column. When the nonzero elements have no phases and are equal to one, it becomes the standard permutation matrix. The complex permutation matrix is mathematically known as a special monomial matrix, and has been used to characterize the mutually unbiased bases \cite{bw09,mb15}. The complex permutation gate is of interest to the study of quantum computation, as it is a somewhat classical part of a quantum circuit; see its use in the definition of the Fourier hierarchy in \cite{Shi2005335}. The diagonal unitary, which is a special complex permutation matrix, can be efficiently simulated in terms of the Clifford+T basis by the algorithm in \cite{wbs14}. We define the \emph{controlled-permutation matrices} to be bipartite controlled unitaries controlled in the computational basis of one system and with the terms on the controlled side being permutation matrices. The \emph{controlled-complex-permutation matrices} are defined similarly.

To study the decomposition of complex permutation matrices, we present a preliminary lemma, which is actually a form of the Hall's marriage theorem \cite{Halltheorem}. Suppose $V=\sum^{d_A}_{j,k=1} \ketbra{j}{k}\ox V_{j,k}$ is a bipartite operator on the space $\cH=\cH_A\ox\cH_B$. We say that $V$ is \emph{absolutely singular} if there are integers $j_1,\cdots,j_s$ and $k_1,\cdots,k_t$ with $s+t>d_A$, such that $V_{j_a,k_b}=0$. The absolute singularity of $V$ is unchanged up to any product permutation operators on the left- and right-hand sides of $V$ (a product permutation operator is of the form $P_A\ox Q_B$, where $P_A$ and $Q_B$ are local permutation operators; in what follows we only need $Q_B$ to be an identity matrix). Hence an absolute singular $V$ is locally equivalent to another bipartite operator whose left-upper $sd_B \times td_B$ submatrix is zero. Evidently an absolutely singular operator is singular, but the converse is not true. We characterize the absolute singularity as follows.
\bl
\label{le:absolutesingular}
$V=\sum^{d_A}_{j,k=1} \ketbra{j}{k}\ox V_{j,k}$ is not absolutely singular if and only if there are $d_A$ distinct integers $k_1,\cdots,k_{d_A}$, such that the blocks $V_{1,k_1},\cdots,V_{d_A,k_{d_A}}$ are all nonzero.
\el
\bpf
We first present a matrix-based proof, and then provide a proof of the equivalence of the lemma to Hall's marriage theorem, which is known to have several different proofs.

\textbf{Matrix-based proof.} The ``if'' part follows from the definition of absolute singularity. Let us prove the assertion in the ``only if'' part. Assume $V$ is not absolutely singular. This assumption and the assertion are both unchanged up to any product permutation operators on the left- and right-hand sides of $V$. We will refer to the $d_B\times d_B$ blocks in $V$ still as $V_{j,k}$ since there is no confusion.
The assertion is trivial for $d_A=1,2$. Next we shall use induction over $d_A$. The induction hypothesis is that the assertion holds when $d_A$ is replaced by $2,\cdots,d_A-1$, and we will prove the assertion for $d_A$.
Since $V$ is not absolutely singular, we may assume that $V_{11}$ is nonzero up to a suitable product permutation operator on the right-hand side of $V$. If the submatrix $X=\sum^{d_A}_{j,k=2} \ketbra{j}{k}\ox V_{j,k}$ is not absolutely singular, then the assertion follows from the induction hypothesis on $X$. Suppose $X$ is absolutely singular. By performing two suitable product permutation operators, respectively, from the left- and right-hand side of $V$, we may assume that $V_{j,k}=0$ where $j=2,\cdots,s$, $k=t+1,\cdots,d_A$, and $d_A\ge s>t\ge1$. Since $V$ is not absolutely singular, we have $s=t+1$. Using a suitable product permutation operator on the left-hand side of $V$, we may assume that $V=\left(
                   \begin{array}{cc}
                     V_1 & 0 \\
                     V_2 & V_3 \\
                     \end{array}
                 \right)$,
where $V_1$ and $V_3$ are, respectively, $(s-1)d_B\times (s-1)d_B$ and $(d_A-s+1)d_B\times (d_A-s+1)d_B$ submatrices. Since $V$ is not absolutely singular, neither are $V_1$ and $V_3$. The hypothesis induction implies that the assertion holds for both $V_1$ and $V_3$.
Hence the assertion holds for $V$.
This completes the proof.

\textbf{Equivalence of the lemma to Hall's marriage theorem.}  We use the combinatorial formulation of Hall's marriage theorem in \cite{Halltheorem}. It involves some given elements, each of which may be in one or more of some given sets. There is a \emph{marriage condition} that says the number of distinct elements contained in $k$ sets is at least $k$, for any integer $k\ge 0$. A \emph{system of distinct representatives} is a set of distinct elements, each of which is in a different set. Hall's marriage theorem says that a system of distinct representatives exists if and only if the marriage condition is satisfied. Let us now describe the equivalence of the current lemma to the above theorem. Take the sets to be the big rows of $V$ labeled by $j$, and the elements to be the big columns labeled by $k$, and let an element $k$ be in a set $j$ if and only if the $V_{j,k}$ is nonzero. Then the marriage condition corresponds to the definition of absolute singularity, and a system of distinct representatives corresponds to a sequence of $d_A$ distinct big column labels $k_i$ ($i=1,\dots,d_A$) such that $V_{i,k_i}$ is nonzero. This establishes the equivalence.
\epf

\bt\label{thm:u_controlled2}
Any bipartite complex permutation unitary has a $3$-sandwich form, composed of controlled-complex-permutation matrices. In particular, if the unitary is a permutation matrix, the $3$-sandwich form is composed of controlled-permutation matrices.
\et
\bpf
The second claim implies the first claim, since any complex permutation unitary is the product of a permutation matrix and a diagonal unitary, the latter can be absorbed into one of the controlled-permutation matrices in the decomposition of the complex permutation unitary. Therefore it suffices to prove the second claim. Suppose $U$ is a bipartite permutation unitary on the $d_A\times d_B$ system.

Let $U=\sum^{d_A}_{j,k=1} \ketbra{j}{k}\ox U_{j,k}$. Since it is not absolutely singular, it follows from Lemma \ref{le:absolutesingular} that there are $d_A$ distinct integers $k_1,\cdots,k_{d_A}$, such that the blocks $U_{1,k_1},\cdots,U_{d_A,k_{d_A}}$ are all nonzero. There are two controlled-permutation matrices $V=\sum^{d_A}_{j=1} \proj{j}\ox V_j$ and $W=\sum^{d_A}_{j=1} \proj{j}\ox W_j$ from the A side, such that the first entry of any one of the blocks $V_1U_{1,k_1}W_{k_1},\cdots,V_{d_A}U_{d_A,k_{d_A}}W_{k_{d_A}}$ of $VUW$ is one. If $d_B=2$ then $VUW$ is a controlled-permutation unitary from the B side. So the assertion holds. We use the induction over $d_B\ge2$.
We have $VUW=X\ox\proj{1}+Y$, where $X$ is a permutation matrix on $\cH_A$, and $Y$ is a permutation matrix on $\cH_A\ox\ket{1}^{\perp}$. The induction hypothesis on $Y$ implies that $Y=Y_1Y_2Y_3$, where $Y_1,Y_2$ and $Y_3$ are controlled-permutation matrices from A, B and A side, respectively. Hence
\bea
U&=V^\dg(I_A\ox\proj{1}+Y_1)(X\ox\proj{1}+Y_2)\notag\\
&\cdot\,(I_A\ox\proj{1}+Y_3)W^\dg,
\eea
which is a 3-sandwich form of $U$ composed of controlled-permutation matrices. So we have proved the second claim.
This completes the proof.
\epf

\smallskip
It is known that the SWAP$_d$ gate defined in Lemma~\ref{le:swap} has a decomposition using three bipartite controlled gates \cite{gc13}. In agreement with the construction in \cite{gc13}, Theorem \ref{thm:u_controlled2} shows that the three gates can be chosen as controlled-permutation gates in a 3-sandwich form.

The theorem also has implications for classical circuits. Define a \emph{classical reversible circuit} (\emph{classical permutation gate}) to be a classical circuit that is a permutation on the allowed set of input data. In the bipartite case, suppose $d_A$ and $d_B$ are the number of possible states on the systems $A$ and $B$, respectively, then we say the circuit acts on a $d_A\times d_B$ system. For example, when $n_A$ and $n_B$ are the number of bits on the two systems, we have $d_A=2^{n_A}$ and $d_B=2^{n_B}$. From Theorem~\ref{thm:u_controlled2}, and noting that in the proof of Theorem~\ref{thm:u_controlled2} there is no requirement of coherence in both the target unitary and the controlled unitaries in the decomposition, we have
\bcr\label{cr:u_classical}
Suppose $T$ is a classical reversible circuit on a $d_A\times d_B$ bipartite system, then $T$ can be implemented using the product of $3$ bipartite classical controlled-permutation gates.
\ecr
Note the classical controlled-permutation gates are controlled in the computational basis, as one would expect.  Corollary~\ref{cr:u_classical} is also stated in Sec.~3.15 and Appendix E of \cite{DeVos10}, where the proof approach is by considering the permutation accomplished by the circuit and directly using the Birkhoff-von Neumann theorem (explained below), which has an integer-arithmetic version that says the following: Any matrix of size $n\times n$ with non-negative integer entries and with row and column sums equal to $q$ can be decomposed as the sum of $q$ permutation matrices of size $n\times n$. Such a statement appears in \cite{Peng06}, and a simple proof is by repeated use of Hall's marriage theorem, each time finding a permutation matrix, which is to be subtracted from the original matrix, and this process terminates when the resulting matrix becomes the zero matrix. The construction of the $3$ classical permutation gates is as follows: arrange the $d_A\times d_B$ computational-basis states in a rectangular table with $d_A$ rows and $d_B$ columns, and define a matrix $M$ to contain integer elements $M_{ij}$ that indicate how many elements in row $i$ of the table are to be transferred to row $j$ of the table after the permutation gate $T$. The first, second, and third controlled-permutation gates permutes among elements in the same row, column, and row, respectively. Each column of the rectangular table after the first gate contains elements that are to be permuted under the second gate. Each permutation in a column corresponds to one permutation matrix in the decomposition of $M$ as the sum of permutation matrices. The argument above roughly describes the proof in \cite{DeVos10} for Corollary~\ref{cr:u_classical}. In comparison, our matrix-based approach for obtaining the circuit decomposition hints at some connections to the sandwich form of general unitaries of the sort in Lemma~\ref{le:2xddecomp} and Theorem~\ref{thm:u_controlled}.

\smallskip

\section{Decomposition of multipartite unitary operators}

\label{sec:multi}

In this section, we study the decomposition of $n$-partite unitary operators $U$ on the space $\ox^n_{j=1}\cH_j$ with $\dim\cH_i=d_i$. We define a \textit{generalized m-sandwich form} of $U$ to be a decomposition of the form $U=U_1U_2\cdots U_m$, where any $U_i$ is a controlled unitary controlled in the computational basis from $n-1$ fixed systems. For example, $U_1$ may be controlled from the systems of $\cH_1,\cdots,\cH_{n-1}$, $U_2$ may be controlled from the systems of $\cH_1,\cdots,\cH_{n-2},\cH_n$, etc. The computational basis in $\ox^n_{j=1}\cH_j$ consists of the product states $\ket{j_1,\cdots,j_n}$ where $j_i=1,\cdots,d_i$ for each $i$. The word ``fixed'' means the choices of controlling parties are fixed for each gate $U_i$.  Such choices are a function of the generalized $m$-sandwich form that we choose. In the results in this section, we always fix such choices. We have

\bpp
\label{pp:multi_controlled}
Any $n$-partite unitary has a generalized $[2\prod^{n-1}_{j=1}(2d_j-2)-1]$-sandwich form.
\epp
\bpf
Let $f(n)=2\prod^{n-1}_{j=1}(2d_j-2)-1$. The assertion is trivial  for $n=1$, and follows from Theorem \ref{thm:u_controlled} for $n=2$. We use the induction on $n$. Assume that any $(n-1)$-partite unitary has a generalized $f(n-1)$-sandwich form.
Let $U$ be an $n$-partite unitary. By regarding $\cH_A=\cH_1$ and $\cH_B=\ox^n_{j=2}\cH_j$ in Theorem \ref{thm:u_controlled}, we obtain the $(4d_1-5)$-sandwich form
\bea
\label{eq:multi_controlled}
U=\prod^{4d_1-5}_{j=1}U_j,
\eea
where $U_j$ is controlled in the computational basis of $\cH_A$ for odd $j$, and of $\cH_B$ for even $j$, respectively. In particular, the computational basis in the latter is realized by performing suitable unitaries on $\cH_B$ that can be absorbed by the $U_j$ with odd $j$. Then
\bea
\label{eq:multi_controlled2}
U_j=\bigoplus^{d_1}_{k=1} \proj{k}\ox U_{jk},  \,\,\forall \mbox{ odd }j,
\eea
where each $U_{jk}$ is a unitary on $\cH_B$. From the induction assumption, $U_{jk}$ has a generalized $[2\prod^{n-1}_{j=2}(2d_j-2)-1]$-sandwich form. Then \eqref{eq:multi_controlled2} implies that $U_j$ with any odd $j$ has a generalized $[2\prod^{n-1}_{j=2}(2d_j-2)-1]$-sandwich form. Since $U_j$ with any even $j$ is a controlled unitary controlled in the computational basis of $\cH_B$, \eqref{eq:multi_controlled} implies that $U$ has a generalized $m$-sandwich form where
\bea\label{eq:multi_controlled3}
m
&=&(2d_1-2)[2\prod^{n-1}_{j=2}(2d_j-2)-1]+2d_1-3
\notag\\
&=&f(n).
\eea
This completes the proof.
\epf

\smallskip
The proof above first divides the systems into two groups of one party and $(n-1)$ parties each. When $n\ge 4$, there are also other ways of dividing the systems at the first step that may give rise to fewer gates in the generalized sandwich form.  The following result is for the case of $n=4$.

\bpp
\label{pp:multi_controlled2}
Any unitary on four parties $A,B,C,D$ has a generalized $[4 (d_A d_B - 1)(2 d_A + 2 d_C -5)- 4 d_A +5]$-sandwich form.
\epp
\bpf
Let $U$ be a unitary on these four parties. By regarding $\cH_A$ and $\cH_B$ in Theorem~\ref{thm:u_controlled} as $\cH_{AB}$ and $\cH_{CD}$ respectively, we obtain the following sandwich form
\bea
\label{eq:multi_controlled4}
U=\prod^{4d_A d_B-5}_{j=1}U_j,
\eea
where $U_j$ is controlled in the computational basis of $\cH_{AB}$ for odd $j$, and in the computational basis of $\cH_{CD}$ for even $j$, respectively.  Then
\bea
\label{eq:multi_controlled5}
U_j=\bigoplus^{d_A d_B}_{k=1} \proj{k}_{AB}\ox U_{jk},  \,\,\forall \mbox{ odd }j,
\eea
where $\proj{k}_{AB}$ are projectors onto the computational basis of $\cH_{AB}$, and each $U_{jk}$ is a unitary on $\cH_{CD}$. From Theorem~\ref{thm:u_controlled}, $U_{jk}$ has a generalized $(4d_C-5)$-sandwich form. Then \eqref{eq:multi_controlled5} implies that $U_j$ with any odd $j$ has a generalized $(4d_C-5)$-sandwich form. Similarly, $U_j$ with any even $j$ has a generalized $(4d_A-5)$-sandwich form. Therefore $U$ is the product of
\bea
\label{eq:multi_controlled6}
&&(2d_Ad_B-2)(4d_C-5)  + (2d_Ad_B-3)(4d_A-5)\notag\\
&=&4 (d_A d_B - 1)(2 d_A + 2 d_C -5)- 4 d_A +5
\eea
unitaries that are controlled in the computational basis of $3$ parties. This completes the proof.
\epf

\smallskip
To compare the two Propositions above, assume $n=4$ in Proposition~\ref{pp:multi_controlled} with the subscripts $1,2,3,4$ replaced by $A,B,C,D$, respectively, and that $d_A\le d_B\le d_C\le d_D$. Then Proposition~\ref{pp:multi_controlled} gives that $U$ is the product of $16(d_A-1)(d_B-1)(d_C-1)-1$ unitaries that are controlled from $3$ parties.  Therefore, at least when $d_B << d_C$ and $d_A$ is a large constant (say $d_A\ge 20$), Proposition~\ref{pp:multi_controlled2} gives a smaller number than Proposition~\ref{pp:multi_controlled}.

The proofs of the results above imply that, if we could reduce the number of bipartite controlled unitaries in the sandwich form in Theorem \ref{thm:u_controlled}, then the number of multipartite controlled unitaries in the generalized sandwich form could also be reduced. In particular, from Theorem.~\ref{thm:u_controlled2}, we have
\bpp\label{pp:multi_controlled3}
Any $n$-partite complex permutation unitary has a generalized $(2n-1)$-sandwich form composed of controlled-complex-permutation unitaries controlled by $n-1$ parties.
\epp
\bpf
It suffices to consider permutation unitaries, for the same reason as stated in the proof of Theorem~\ref{thm:u_controlled2}. From Theorem~\ref{thm:u_controlled2}, the claim holds for $n=2$. The proof is by induction over $n$. The induction hypothesis is that the claim holds when $n$ is replaced by any positive integer less than $n$. Now consider $n\ge 3$, and take a bipartite cut of the first $n-1$ parties versus the last party. From Theorem~\ref{thm:u_controlled2}, the permutation unitary has a $3$-sandwich form, and the first and the last gates in the $3$-sandwich form are a controlled permutation controlled from the first $n-1$ parties. The middle gate in the $3$-sandwich form is a controlled permutation controlled from the last party, so it is of the form $U_1\ox \ketbra{1}{1} + U_2 \ketbra{2}{2}$, where the permutation operators $U_1$ and $U_2$ on the first $n-1$ parties can each be decomposed into $2(n-1)-1$ controlled-permutation gates controlled by $n-2$ parties, and the choices of those controlling $n-2$ parties are always the same for the decompositions of $U_1$ and $U_2$, according to the induction hypothesis. Therefore the permutation unitary on $n$ parties has a generalized $(2n-1)$-sandwich form composed of controlled-permutation gates controlled by $n-1$ parties. The case with phases is similar, just adding the word ``complex''. This completes the proof.
\epf

The result above has a corresponding statement for classical reversible circuits. In the special case that each party is one bit, it is illustrated by a sample circuit in Fig. 2 of \cite{DeVos10b} (note the sequence of lines is opposite from that in the proof above).

\smallskip
As mentioned in Sec.~\ref{sec:bipartite}, it is possible that the literature results on the decomposition of qudit circuits \cite{blb05,bbo06,dw13} could be combined with the results in this section to give better upper bounds of the number of two-qudit gates.

\section{Decomposition using a simple type of gates}

\label{sec:elementary}

In this section, we apply our result on decomposition using controlled unitaries to the decomposition using more basic type of gates defined below.  One of our motivations is to characterize the nonlocal part of the cost for implementing bipartite unitaries using some measure with a fixed unit, rather than using the number of controlled unitaries which is a measure with its unit dependent on the dimensions.   The cost measure that we use is the number of standard gates defined below, and we do not allow any ancillary systems in the circuit. The case with ancillas will be discussed in Sec.~\ref{sec:ancilla}. In the following definitions, $I_X$ stands for the identity operator on system $X$.

\bd\label{def:ele1}
A standard gate is a unitary acting on the Hilbert space $\cH_{AB}=\cH_A\ox\cH_B$ of the form $U=U_{AB}=(V_{ab}\oplus I_{AB\backslash ab})$, where $\cH_{AB}=\cH_{ab}\oplus \cH_{AB\backslash ab}$, and $\cH_a\subseteq \cH_A$ and $\cH_b\subseteq \cH_B$ are two-dimensional each, and $V_{ab}$ is a Schmidt-rank-2 unitary on the $2\times 2$ space $\cH_{ab}=\cH_a \ox \cH_b$. The $V_{ab}$ is called the nontrivial part of $U$.
\ed

Note the word ``Schmidt-rank-2'' above can be replaced by ``controlled'', as Schmidt-rank-2 unitaries are controlled unitaries (\cite{cy13}; also see an alternative proof in \cite{cy14ap}), and two-qubit unitaries of Schmidt rank greater than 2 must have Schmidt rank 4 \cite{Nielsen03} and thus cannot be controlled unitaries. The case of $\cH_{AB}$ being strictly larger than $\cH_{ab}$ is useful, for example, in the decomposition of the Toffoli gate \cite{rrg07}, and has been experimentally realized \cite{Lanyon09}. The definition above can be extended to a more general definition below:

\bd\label{def:ele2}
A bipartite elementary gate is a unitary acting on the Hilbert space $\cH_A\ox\cH_B=(\cH_{a}\ox \cH_C \oplus \cH_D)\ox(\cH_{b}\ox \cH_E \oplus \cH_F)$ of the form $U=(V_{ab}\ox I_{CE}) \oplus I_{AB\backslash abCE}$, where $\cH_a$ and $\cH_b$ are two-dimensional each, and $V_{ab}$ is a Schmidt-rank-2 unitary, and $\cH_{AB}=\cH_{AB\backslash abCE}\oplus\cH_{abCE}$.
\ed

In the following we consider the decomposition of bipartite unitary operators into the product of bipartite standard gates defined in Definition~\ref{def:ele1} and arbitrary local gates, with the goal of minimizing the number of nonlocal standard gates. The more general Definition~\ref{def:ele2} will not be studied in this paper except that we define some gate cost using it in Definition~\ref{def:ele4} and raise some open questions.

We define the following gate-cost measures for a bipartite unitary.
\bd\label{def:ele4}
Let $\cH=\cH_A\otimes\cH_B$ be the complex Hilbert space of a finite-dimensional bipartite quantum system, with $\dim\cH_A=d_A$ and $\dim\cH_B=d_B$. For any given bipartite unitary $U: \cH\ra\cH$,
\begin{eqnarray}
\label{eq:cu}
c_s(U) &:= \min\{k | U=U_1U_2\cdots U_k,~~U_i\in \cS_{s}\},\notag\\
c_e(U) &:= \min\{k | U=U_1U_2\cdots U_k,~~U_i\in \cS_{e}\},
\end{eqnarray}
where $\cS_{s}$ (respectively, $\cS_{e}$) is the set of bipartite unitaries on the same space that are equivalent to the standard (respectively, bipartite elementary) gates under local unitaries.
\ed

%It is natural to ask whether the quantities defined above are unchanged upon adding some ancillary subspaces or subsystems that undergo trivial dynamics. These questions are listed in the Conclusions section.

In the case $d_A=d_B=2$, it is well known that three Schmidt-rank-2 gates are sufficient and necessary for a general two-qubit unitary \cite{vw04}, as mentioned in Sec.~\ref{sec:bipartite}. An example that needs three Schmidt-rank-2 gates is the two-qubit SWAP gate (\cite{vw04}, also see Lemma~\ref{le:swap}).  Our main result for general $d_A\times d_B$ system is as follows.

\bpp
\label{pp:elem}
(i) Any bipartite unitary on $d_A\times d_B$ system is the product of $f(d_A,d_B)$ standard gates interspersed with local unitaries on $\cH_A$ or $\cH_B$, where
\bea
\label{eq:elem1}
f(d_A,d_B)
&=&2(d_A-1)^2 \lfloor \frac{d_B}{2}\rfloor
\notag\\
&+&(2d_A-3)(d_B-1) \lfloor \frac{d_A}{2}\rfloor.
\eea

(ii) If the unitary is a controlled unitary controlled from the $A$ side, then
\bea
\label{eq:elem1b}
f(d_A,d_B)=(d_A-1)\lfloor\frac{d_B}{2}\rfloor.
\eea

(iii) If the unitary is a complex permutation unitary, then
\bea
\label{eq:elem1c}
f(d_A,d_B)=2(d_A-1)\lfloor\frac{d_B}{2}\rfloor+(d_B-1)\lfloor\frac{d_A}{2}\rfloor.
\eea
(iv) If the nontrivial part of the standard gates is required to be CNOT, then at most $3(d_A-1)(d_B-1)$ such standard gates
together with local permutation gates can implement any bipartite permutation unitary on $d_A\times d_B$ space.
\epp

\bpf
(i). Let $U$ be the bipartite unitary. Theorem~\ref{thm:u_controlled} implies that $U$ has the following sandwich form
\bea
\label{eq:elem2}
U=\prod^{4d_A-5}_{j=1}U_j,
\eea
where $U_j$ is controlled in the computational basis of $\cH_A$ for odd $j$, and in the computational basis of $\cH_B$ for even $j$, respectively.  For all odd $j$, we have
\bea
\label{eq:elem3}
U_j&=&\bigoplus^{d_A}_{k=1} \proj{k}_A\ox U_{jk},\notag\\
&=&\prod^{d_A}_{k=1} [\proj{k}_A\ox U_{jk} \oplus (I_A-\proj{k}_A)\ox I_B ],\quad
\eea
where $\proj{k}_A$ are projectors onto the computational basis of $\cH_A$, and each $U_{jk}$ is a unitary on $\cH_B$.
We can apply a local unitary $U_{jd_B}$ on $\cH_B$ before performing other steps below. In order to implement $U_j$, the operator that remains to be implemented
is still given by \eqref{eq:elem3} but with $U_{jd_B}$ becoming the identity matrix, and the other operators $U_{jk}$ also  changed but we
still denote the changed matrices as $U_{jk}$, with $1\le k\le d_A-1$.  The $U_j$ is to be implemented using the product
of $d_A-1$ operators, as shown in the second line of \eqref{eq:elem3}. Then each of the $U_{jk}$ with $1\le k\le d_A-1$ can be assumed to be a diagonal unitary, because we can apply a suitable local unitary similarity transform on $\cH_B$ so that $U_{jk}$ is diagonal and $I_B$ is unchanged. By a local diagonal unitary gate on $\cH_A$ which only applies a phase on $\ket{k}_A$, we can set the last diagonal element
of $U_{jk}$ to be $1$, while the $I_B$ corresponding to the basis kets in $\cH_A$ other than $\ket{k}_A$ are unchanged.
Therefore we have
\bea
\label{eq:elem4}
U_{jk}=\diag(x^{(jk)}_1,x^{(jk)}_2,\dots,x^{(jk)}_{d_B-1},1),
\eea
where $x^{(jk)}_i$ are complex phases, $i=1,2,\dots,d_B-1$. Then we choose $\lfloor \frac{d_B}{2}\rfloor$ standard gates as follows:
\bea
\label{eq:elem5}
V^{(jk)}_r=I_A \ox I_B + (x^{(jk)}_{2r-1}-1) \proj{k}_A\ox\proj{2r-1}_B  \notag\\
+ (x^{(jk)}_{2r}-1) \proj{k}_A\ox\proj{2r}_B, \mbox{ for }\,1\le r\le\lfloor \frac{d_B}{2}\rfloor,
~~~~~
\eea
Each gate $V^{(jk)}_r$ applies phases on the two states $\ket{k}_A \ox \ket{2r-1}_B$ and $\ket{k}_A \ox \ket{2r}_B$, but keeps other computational basis states of $\cH_{AB}$ unchanged.
It is easy to verify that such a gate has Schmidt rank at most 2 when viewed as a unitary acting on the $2\times 2$ system with basis $\{\ket{k'}_A,\ket{k}_A\}\times\{\ket{2r-1}_B,\ket{2r}_B\}$, where $k'\neq k$.
Hence for each $(j,k)$ pair with odd $j$ and $1\le k\le d_A-1$ , we need $\lfloor \frac{d_B}{2}\rfloor$ standard gates to implement the operator $\proj{k}_A\ox U_{jk} \oplus (I_A-\proj{k}_A)\ox I_B $ in the last line of \eqref{eq:elem3}. Therefore, for each odd $j$, $U_j$ needs $(d_A-1) \lfloor \frac{d_B}{2}\rfloor$ standard gates to implement, assisted by local unitaries.  Similarly, for each even $j$, $U_j$ needs $(d_B-1) \lfloor \frac{d_A}{2}\rfloor$ standard gates to implement, assisted by local unitaries. The assertion then follows by counting the numbers of $U_j$ in \eqref{eq:elem2} in terms of odd and even $j$. This completes the proof of (i).

(ii). The claim follows from the proof of (i) by setting the upper bound for $j$ in \eqref{eq:elem2} to $1$.

(iii). The claim follows from Theorem~\ref{thm:u_controlled2} and the result of (ii) applied to the $A$ and $B$ sides.

(iv). From Theorem~\ref{thm:u_controlled2} , every bipartite permutation unitary is the product of $3$ controlled-permutation unitaries, controlled from the $A$, $B$ and $A$ side, respectively. Every permutation on $n$ elements is the product of at most $n-1$ transpositions (swap of two elements). Define a controlled-transposition gate to be a bipartite unitary of the form $\ketbra{1}{1}_A\ox I_B + \ketbra{2}{2}_A\ox V_B$, where $V_B=\ketbra{j}{k}+\ketbra{k}{j}+\sum_{i\ne j,k}\ketbra{i}{i}$, for some $j\ne k$ ($\{\ket{i}\}$ is the computational basis of $\cH_B$). For the special case $d_A=2$, up to a local permutation on $\cH_B$ we can write a controlled-permutation gate from the $A$ side as $\ketbra{1}{1}\ox I_B + \ketbra{2}{2}\ox P_2$, where $P_2$ is a permutation unitary on $\cH_B$. This controlled-permutation gate can be written as the product of at most $d_B-1$ controlled-transposition gates, which are standard gates with their nontrivial part being the CNOT. For larger $d_A$, up to a local permutation on $\cH_B$ we can write the controlled-permutation gate as $\ketbra{1}{1}\ox I_B + \sum_j \ketbra{j}{j}\ox P_j$, where $P_j$ are permutation unitaries on $\cH_B$. Take the subspace span$\{\ket{1}_A,\ket{j}_A\}$ ($2\le j\le d_A$) as the $A$ side space in the $d_A=2$ result above; we have that at most $d_B-1$ standard gates with the nontrivial part being CNOT can implement $(I_A-\ketbra{j}{j})\ox I_B + \ketbra{j}{j}\ox P_j$. Repeat this $d_A-1$ times for $j=2,\dots,d_A$, a controlled-permutation gate from the $A$ side can be implemented using at most $(d_A-1)(d_B-1)$ such standard gates. The last result is the same for the $B$ side. Hence the claim follows.
\epf

\smallskip
It can be verified that in Proposition~\ref{pp:elem}(iv), the phrase ``together with local permutation gates'' can be dropped by allowing the nonlocal unitary to be implemented up to local permutations before and after it. Since a permutation gate on $d$-dimensional space requires at most $d-1$ transpositions of the type $\ketbra{j}{k}+\ketbra{k}{j}+\sum_{i\ne j,k}\ketbra{i}{i}$, the four local permutations on $\cH_A$ or $\cH_B$ require at most $2d_A+2d_B-4$ local transpositions in total. Therefore the total number of standard gates of the CNOT type and the local transpositions is at most $3(d_A-1)(d_B-1)+2d_A+2d_B-4=3d_A d_B -d_A -d_B - 1$. It could potentially be further reduced by a constant factor, and this is listed as an open problem in the Conclusions.

From \cite{gc13} and Proposition \ref{pp:elem} (ii), the SWAP$_d$ gate has a decomposition using $3(d-1)\lfloor \frac{d}{2}\rfloor$ standard gates across the two systems, together with some local unitaries.
On the other hand, if we are not restricted to writing the SWAP$_d$ gate as a product of some gates, but consider the actual cost of implementation, we could also make use of tensor products.
Suppose $d=\prod_{j=1}^m p_j$, where $m\ge 1$ is an integer and $p_j$ are primes. Then the SWAP$_d$ gate is the tensor product of the SWAP gates on $p_j\times p_j$ systems. The SWAP gate on $p_j\times p_j$ system has a decomposition using $3(p_j-1)\lfloor \frac{p_j}{2}\rfloor$ bipartite standard gates, together with some local unitaries. Hence the total implementation cost is $\sum_{j=1}^m 3(p_j-1)\lfloor \frac{p_j}{2}\rfloor$ bipartite standard gates, together with some local unitaries.

\section{The role of Schmidt rank in decomposition of bipartite unitaries}
\label{sec:Schmidt}

The Schmidt rank of a bipartite unitary $U$ sometimes determines the number of bipartite controlled unitaries needed to decompose $U$, as it is proved in \cite{cy13,cy14ap} that $c(U)=1$ when ${\rm Sch}(U)=2$ or $3$. To investigate the relation between $c(U)$ and ${\rm Sch}(U)$ for general bipartite unitary $U$, we discuss the different cases characterized by how large $r:={\rm Sch}(U)$ is compared to the dimensions $d_A$ and $d_B$. If $r\ge \min\{d_A,d_B\}$, then it follows from Theorem~\ref{thm:u_controlled} (applied to the $A$ or $B$ side) that $c(U)\le 4r-5$. On the other hand if $r< \min\{d_A,d_B\}$, then we need to count the number of parameters in $U$. It is equal to $(d_A^2-r+d_B^2)r$, which is smaller than $2d_Ad_B^2$ when $d_A\le d_B$. A controlled unitary from the $A$ side contains $d_A d_B^2$ parameters, and noting that there are some redundant parameters when counting consecutive controlled unitaries in a product, theoretically $U$ could be the product of only three controlled unitaries (or even two when $r$ is further restricted to smaller values). But the actual number may be higher. A possible class of candidate examples that \emph{may} need more than three controlled unitaries is the $U'$ in Example~\ref{ex:u_6gates} below.

We now show a class of examples where $c(U)$ is much smaller than ${\rm Sch}(U)$ (note that a generic permutation matrix already has this property, according to Theorem~\ref{thm:u_controlled2}, but our interest here is to show the derived class of examples $U'$ that fit into the requirement $r< \min\{d_A,d_B\}$ in the previous paragraph).
\bex\label{ex:u_6gates}
{\rm
Let $V_{CB}$ be a generic unitary on $d\times d$ system of Schmidt rank $d^2$ with $d>2$, and let $U=V_{CB}\ox I_D$, where $D$ is of the same size as $B$ and $C$ ($d$ dimensions). Then $U$ is of Schmidt rank $d^2$ across the bipartite cut $CD$-$B$. But there is a decomposition using only $6$ controlled unitaries: first, swap the states of the systems $D$ and $B$, using $3$ controlled gates \cite{gc13}, then do the $V$ on $CD$, and finally swap the $D$ and $B$ again, using another $3$ controlled gates. The local unitary on $CD$ in the second step could be absorbed into the two controlled unitaries before and after it, thus only $6$ controlled unitaries are needed in total without extra local unitaries. Now consider the unitary $U'=U \ox I_E\ox I_F$, where $E$ is one qubit and $F$ is of dimension $2d$. Then $U'$ of Schmidt rank $d^2$ across the $CDE$-$BF$ cut, and $c(U')\le 6$, and the Schmidt rank $r=d^2$ satisfies $r< \min\{d_{CDE},d_{BF}\}$, fitting into the requirement in the first paragraph of this section.
}
\eex

Speaking about the general dependence of $c(U)$ on ${\rm Sch}(U)$, the two classes of examples $U$ and $U'$ in Example~\ref{ex:u_6gates} show that $c(U)$ is not lower bounded by a function of ${\rm Sch}(U)$ with maximum or supremum value greater than $6$.  Whether $c(U)$ is upper bounded by a function of ${\rm Sch}(U)$ is unknown, and this is listed as an open question in Sec. \ref{sec:con}.

\subsection{Nonlocal cost of bipartite permutation operators}\label{sec:permutation}

We have shown in Theorem~\ref{thm:u_controlled2} that every bipartite permutation operator can be implemented by three controlled unitaries, but such controlled unitaries may be hard to implement since they are on $d_A\times d_B$ space.  A better measure of the nonlocal part of the gate cost (i.e., the nonlocal gate cost) is in terms of the bipartite elementary gates of Definition~\ref{def:ele2}. Two results that depend on $d_A$ and $d_B$ are given in Proposition~\ref{pp:elem}(iii)(iv), though special classes of the bipartite elementary gates are used therein. In this subsection we study the nonlocal gate cost as a function of the Schmidt rank or dimension of the bipartite permutation unitary.  The obtained upper bounds could be much less than those in Proposition~\ref{pp:elem}(iii)(iv) for some classes of permutation unitaries. The result can also be stated in terms of the entanglement cost under local operations and classical communications (LOCC). Hence it provides a significant class of examples that the entanglement cost of a bipartite unitary is upper bounded by a function of Schmidt rank independent of the dimensions. The only known result of this flavor is about Schmidt-rank-$2$ unitaries, which are implementable using one ebit of entanglement under LOCC \cite{cy13}.
To study the nonlocal gate cost, we first present some definitions and preliminary lemmas.\bd\label{def:pp}
A partial permutation matrix is a matrix with elements $0$ or $1$ only, with at most one nonzero element on each row and column.  The input (respectively, output) space for such matrix is the complex Hilbert space that is the span of the computational basis states corresponding to the nonzero columns (respectively, rows) of the matrix.
\ed

\bd\label{def:pprank}
The partial-permutation rank of a bipartite operator $U$, denoted ${\rm ppr}(U)$, is the minimum number of terms $q$ such that
\bea\label{eq:ppexpansion}
U=\sum_{j=1}^q A_j \ox B_j,
\eea
where $A_j$ and $B_j$ are partial permutation operators on $\cH_A$ and $\cH_B$.
\ed
The above two definitions imply that if a bipartite operator has a partial-permutation rank, then its entries are non-negative integers. So the partial-permutation rank is not defined for a bipartite operator containing a negative or non-integer entry in its matrix. The partial-permutation rank of the bipartite permutation matrices will be studied in Lemma~\ref{le:partial_permutation2}.

\bl\label{le:bipartite_CNOT}
Suppose $U$ is a bipartite controlled unitary of the form $P_1\ox V_1 + P_2 \ox V_2$, where $P_1$ and $P_2$ are orthogonal projectors on $\cH_A$, and $V_1$ and $V_2$ are unitaries on $\cH_B$. With the help of a one-qubit ancilla on each side, $U$ can be implemented using two bipartite CNOT gates and some local unitary gates. The initial and final states of each ancilla qubit are the same.
\el
\bpf
Let $a$ and $b$ denote the qubit ancillas on the $A$ and $B$ side initialized in the state $\ket{0}_a$ and $\ket{0}_b$, respectively.
The unitary $U$ can be implemented using the ancillas and one CNOT gate with the following sequence of gates: a controlled gate on $\cH_{Aa}$: $V_{Aa}=P_1 \ox I_a+ P_2\ox X_a$, where $X_a=\ketbra{0}{1}_a+\ketbra{1}{0}_a$ (similar below with subscripts changed), and a CNOT gate on $\cH_{ab}$: $\mbox{CNOT}_{ab}=\ketbra{0}{0}\ox I_b + \ketbra{1}{1}\ox X_b$, and a controlled gate on $\cH_{bB}$: $W_{bB}=\ketbra{0}{0}\ox V_1 + \ketbra{1}{1}\ox V_2$, and then $\mbox{CNOT}_{ab}$ again to erase the state on $b$ to $\ket{0}_b$, and the $V_{Aa}$ again to erase the state on $a$ to $\ket{0}_a$. This implements $U$ without changing the states of $a$ and $b$.
\epf

\bl\label{le:partial_permutation}
Suppose $U$ is a bipartite permutation unitary of partial-permutation rank $q$. Then the following statements hold:\\
(i) with the help of a one-qubit ancilla on one party and a two-qubit ancilla on the other party, $U$ can be implemented using at most $6q$ bipartite CNOT gates and some local permutation gates.\\
(ii) with the help of a one-qubit ancilla on either party and $3q$ ebits of entanglement, $U$ can be implemented using LOCC.
\el
\bpf
(i) Consider the matrix representation of $U$ in the computational basis of $\cH_A\ox \cH_B$.   Then $U$ can be expanded as $U=\sum_{j=1}^q A_j\ox B_j$, where $A_j$ and $B_j$ are partial permutation matrices.
Each $A_j$ or $B_j$ has an input space and an output space as defined in Definition~\ref{def:pp}. Assume without loss of generality that the two-qubit ancilla is on the $B$ side. Denote the ancilla qubit on the $A$ side as $a$, and the two ancilla qubits on the $B$ side as $b$ and $c$. Let $\{\ket{0},\ket{1}\}$ (with suitable subscripts $a$,$b$,$c$) be the computational basis of each ancilla qubit. Let the initial state of each of the ancilla qubits be $\ket{0}$.

Now let us visualize the computational basis of $\cH_A\ox \cH_B$ as a rectangular table, with the rows labeling the computational basis states of $\cH_A$, and columns for those of $\cH_B$.  The input spaces of $A_j\ox B_j$ ($j=1,\dots,q$) correspond to small (disconnected) rectangles in such a table. In other words, the computational basis states in the input space of $A_j\ox B_j$ take all intersections of some rows and some columns of the table. In the following we abbreviate the word ``disconnected'' since it turns out that the rows and columns in such a ``small rectangle'' need not be consecutive in our argument.

The whole table of size $d_A\times d_B$ is thus partitioned into $q$ disjoint small rectangles.  The output spaces of $A_j\ox B_j$ also correspond to small rectangles in the table. Our goal is to move the small rectangles to their respective desired positions, while for each such rectangle, we also hope to do an internal permutation of elements according to the form of $A_j$ and $B_j$.  Such internal permutation of elements is the tensor product of two permutations on a subspace of $\cH_A$ and a subspace of $\cH_B$, respectively. But given that there may be some overlap between the input rectangle for one $j$ and the output rectangle for a different $j$, it is hard to do an in-place swap of the rectangles. We avoid this problem by making use of the ancilla qubit $c$, since it effectively supplies two copies of the whole table of size $d_A\times d_B$, corresponding to the states $\ket{0}_c$ and $\ket{1}_c$, respectively. The latter copy is called the \emph{backup copy} below.

For each $j=1,\dots,q$, we perform the following procedure which consists of $3$ controlled-permutation gates. Denote by $P$ the rectangle corresponding to the input space of $A_j\ox B_j$ in the original copy of the table. We first do a controlled-permutation unitary controlled in the computational basis of $\cH_A$ to swap the elements in $P$ into the place (denoted by $M$) in the backup copy of the table and in the target columns. Then perform a controlled-permutation unitary controlled in the computational basis of $\cH_B\ox \cH_c$ to swap the block $M$ into the desired rows in the backup copy (denote the target rectangle by $Q$). Now the part of $M$ that is not in $Q$ (denoted by $M\backslash Q$) is an all-zero block (for any input state of the form $\ket{\psi}_{AB}\ox \ket{0}_c$), but it should have the original contents before these two gates were applied, as it is not the output position for the original $P$. Therefore, we lastly perform a controlled unitary controlled in the computational basis of $\cH_A$ to swap the partial rectangle $M\backslash Q$ and its corresponding part in $P$.  Note that if $M\backslash Q$ is an empty set, the last two unitary gates are actually the identity operation. After the $3$ gates, the original probability amplitudes in $M\backslash Q$ are unchanged, but the probability amplitudes in $P$ and $Q$ are swapped. Since the original state had zero probability amplitude in $Q$ in the backup copy (note this is still true for $j>1$ according to our procedure here), the state after these $3$ gates has zero probability amplitude in the rectangle $P$ (in the original copy).  The internal permutations required in each rectangular block can be accomplished in the first two of these three controlled-permutation gates.

After performing the (at most) $3q$ controlled-permutation gates, we do a local $X_c$ gate on particle $c$ to swap the states $\ket{0}_c$ and $\ket{1}_c$. Now the $U$ is implemented and the ancilla qubits are back in their original state. The local gate $X_c$ can be absorbed into the last one of those (at most) $3q$ gates, which is a bipartite controlled-permutation gate controlled in the computational basis of $\cH_A$. Thus $U$ is implemented using at most $3q$ controlled-permutation unitaries, with the help of the ancilla qubit $c$. Lemma~\ref{le:bipartite_CNOT} implies that each of these controlled-permutation gates, which can be written in two terms, can be implemented using two bipartite CNOT gates and some local gates, the latter are local permutation gates in the current case. The two ancilla qubits used are $a$ and $b$, and they can be recycled through these applications of Lemma~\ref{le:bipartite_CNOT}, because they start and end in the $\ket{0}$ state in each application. Hence at most $6q$ bipartite CNOT gates and some local permutation gates can implement $U$ with the help of the ancilla qubits $a$, $b$ and $c$. This completes the proof of (i).

(ii) The proof is similar to (i), but note that each of the (at most) $3q$ bipartite controlled-permutation gate with two terms can be implemented using 1 ebit of entanglement and LOCC \cite{ygc10}. Thus we need at most $3q$ ebits in total and also need the ancilla qubit $c$, but do not need the ancillary qubits $a$ and $b$. This completes the proof of (ii).
\epf

\smallskip
Next we relate the partial-permutation rank with the Schmidt rank.
\bl\label{le:partial_permutation2}
Suppose the bipartite permutation unitary has partial-permutation rank $q$ and Schmidt rank $r$. Then $q\le\min\{d_A^2, d_B^2, d_A r, d_B r, 2^r\}$.
\el
\bpf
Suppose $U$ is the bipartite permutation unitary on the $d_A\times d_B$ system. The matrix $U$ consists of $d_A^2$ blocks, each of which is a $d_B\times d_B$ partial permutation matrix. We denote them by $B_{jk}$, so that $U=\sum_{jk} \ketbra{j}{k}\ox B_{jk}$.
Since $\ketbra{j}{k}$ is also a partial permutation matrix, we have $q\le d_A^2$ and by symmetry $q\le d_B^2$.
Since $U$ is a permutation matrix, the nonzero blocks $B_{jk}$ for fixed $j$ are linearly independent. So the number of them is not greater than $r$. It holds for $j=1,2,\dots,d_A$. Thus the total number of nonzero $B_{jk}$ is not greater than $d_A r$. Therefore $q\le d_A r$. By symmetry of the $A$ and $B$ sides, we have $q\le d_B r$.

It remains to prove $q\le 2^r$. Suppose $\{F_i\}_{i=1}^r$ is a set of $r$ linearly independent blocks among $B_{jk}$.  All other blocks that are not included in the set $\{F_i\}_{i=1}^r$ are linear combinations of the $F_j$. This last property does not change if we replace $\{F_i\}_{i=1}^r$ by $\{G_i\}_{i=1}^r$. Here each $G_i$ is a linear combination of the $F_j$, and is of the standard form $G_i(t)=\delta_{it}$, $i=1,2,\dots,r$, where $G_i(t)$ is the $t$-th matrix element of $G_i$ according to some fixed ordering of the matrix elements, and $\delta_{it}$ is the Kronecker delta function. Note that such orderings of matrix elements must exist but may not be the usual row-first ordering, as it depends on the operators $F_i$. Then any $B_{jk}$ as a linear combination of $G_i$ ($i=1,2,\dots,r$) must satisfy that the coefficient for $G_i$ is either $0$ or $1$, since the resulting matrix is a $(0,1)$-matrix which implies that its first $r$ elements (in the ordering above) must be either $0$ or $1$. Thus there are at most $2^r$ choices of the ordered set of coefficients, leading to at most $2^r$ distinct blocks $B_{jk}$. Denote the distinct $B_{jk}$ as $D_l$, $l=1,2,\dots,m$. Then $m\le 2^r$, and $U=\sum_{l=1}^m \left(\sum_{(j,k)\in S_l}\ketbra{j}{k}\right)\ox D_l$, where $S_l=\{(j,k): B_{jk}=D_l\}$. Since $U$ is a permutation matrix, the operators $\sum_{(j,k)\in S_l}\ketbra{j}{k}$ are partial permutations, for any $l$. Hence $q\le m\le 2^r$. This completes the proof.
\epf

\smallskip
Lemmas~\ref{le:partial_permutation} and \ref{le:partial_permutation2} immediately imply

\bt\label{thm:u_permutation}
Suppose $U$ is a bipartite permutation unitary of Schmidt rank $r$. With the help of a one-qubit ancilla on one party and a two-qubit ancilla on the other party, $U$ can be implemented using at most $6\min\{d_A^2, d_B^2, d_A r, d_B r, 2^r\}$ bipartite CNOT gates and some local permutation gates. Alternatively, $U$ can be implemented using LOCC and at most $3\min\{d_A^2, d_B^2, d_A r, d_B r, 2^r\}$ ebits of entanglement with the help of one ancillary qubit on either party.
\et

The theorem can also be stated for classical reversible circuits, by making minor changes such as replacing ``qubit'' with ``bit'', and ``local permutation gates'' with ``local reversible gates''.

In Lemma~\ref{le:partial_permutation2}, the dimension-independent upper bound $2^r$ may still be improved.
However, the following Example~\ref{ex:example1} provides evidence that at least for a class of bipartite permutation matrices, the partial-permutation rank grows fast with $r$ (but is not known to be exponential). The operational meaning of the Schmidt rank $r$ is that its logarithm is an upper bound of how many ebits of entanglement a bipartite unitary can create starting from a product state (possibly with local ancillas). Thus the separation between the partial-permutation rank and the Schmidt rank gives some indication about the separation of the entangling power and the entanglement cost under our protocol in the proof of Lemma~\ref{le:partial_permutation}.

Before presenting the example, we first define some versions of ranks for matrices. Let $\rank(T)$ denote the usual rank of a matrix $T$.

\bd\label{def:ranks}
(i). For a matrix $T$ with nonnegative elements, the nonnegative rank \cite{nonnegativerank} $\rank^{+}(T)$ is the minimum number of rank-1 matrices with nonnegative elements that sum to $T$.\\
(ii). For a binary matrix $T$ (binary means the elements are 0 or 1), the binary rank \cite{xorrank} $\rank_N(T)$ is the minimum number of rank-1 binary matrices that sum to $T$.\\
(iii). For a binary matrix $T$, the XOR rank \cite{xorrank} $\rank_X(T)$ (also called modulo-2 rank) is the minimum number of rank-1 binary matrices such that their sum modulo 2 is $T$. It is also equal to the rank over the finite filed $\mathbb{F}_2$, or the number of linearly independent rows (or columns) under arithmetic operations in $\mathbb{F}_2$.
\ed
It is apparent that $\rank(T)\le\rank^{+}(T)\le\rank_N(T)$ and $\rank_X(T)\le\rank_N(T)$ hold for any binary matrix $T$, and according to \cite{xorrank}, $\rank_X(T)$ and $\rank(T)$ are generally incomparable.

\bex\label{ex:example1}
{\rm
Consider bipartite permutation unitaries of the form
\bea\label{eq:perm_example}
U=\sum_{i=1}^M (\ketbra{2i-1}{2i-1} + \ketbra{2i}{2i})\ox  (I_B-C_i) \notag\\
+ (\ketbra{2i-1}{2i} + \ketbra{2i}{2i-1})\ox  C_i,
\eea
where $C_i$ are $d_B\times d_B$ diagonal partial permutation matrices, $i=1,2,\dots,M$. The diagonal part of $U$ is $U_{diag}=\sum_{i=1}^M (\ketbra{2i-1}{2i-1} + \ketbra{2i}{2i})\ox  (I_B-C_i)$. It is a partial permutation matrix, but its elements are all diagonal so the implementation is trivial, thus the implementation cost of $U$ in terms of the protocol in the proof of Lemma~\ref{le:partial_permutation} is determined by the off-diagonal part of $U$, which is denoted $U_{od}:=\sum_{i=1}^M (\ketbra{2i-1}{2i} + \ketbra{2i}{2i-1})\ox  C_i$. Hence ${\rm ppr}(U_{od})$ is proportional to the nonlocal implementation cost under our protocol.  The diagonal elements of the matrices $C_i$ can be rearranged into a matrix $T$ of size $M\times d_B$ with elements $T_{jk}=\bra{k}_B C_j\ket{k}_B$. It is known that $\rank^{+}(T)\le\rank_N(T)$. And $\rank_N(T)={\rm ppr}(U_{od})$, since the minimum-term expansion of $U_{od}$ of the form \eqref{eq:ppexpansion} must involve local operators on $\cH_A$ which are the tensor product of a diagonal partial permutation operator on an $M$-dimensional space and the operator $\ketbra{1}{2}+\ketbra{2}{1}$ on a two-dimensional space, and the partial permutation operators on $B$ are diagonal. These two types of diagonal partial permutation operators mentioned above correspond to the column and row vectors in an expansion of $T$ in terms of direct products of binary column and row vectors. Therefore $\rank^{+}(T)\le\rank_N(T) = {\rm ppr}(U_{od})$. On the other hand, $\rank(T)\ge {\rm Sch}(U_{od})$, since the expansion of $T$ using $\rank(T)$ terms which are the direct products of a column vector and a row vector, corresponds to an expansion of $U_{od}$ in terms of tensor-product operators. Therefore, any separation between $\rank(T)$ and $\rank^{+}(T)$ provides a lower bound for the separation between ${\rm Sch}(U_{od})$ and ${\rm ppr}(U_{od})$. By the way, for this $U$ we have $\vert{\rm Sch}(U_{od})-{\rm Sch}(U)\vert \le 1$, since the diagonal blocks are $I_B-C_i$, and the Schmidt rank is equal to the number of linear independent $d_B\times d_B$ blocks in the matrix $U$. The operational meaning of ${\rm Sch}(U)$ is mentioned before the example.

From \cite{Kol14}, there is a class of $(0,1)$-matrices $T$ such that the separation between $\rank_{\e}(T)$ and of $\rank^{+}_{\e}(T)$ is at least quasipolynomial (but not known to be exponential), more precisely, $\log \rank^{+}_{\e}(T) \ge \Omega\left(\log^2 \rank_{\e}(T)\right)$ for such $T$.  The subscript $\e$ means the $T$ could be replaced by a matrix that approximates $T$ to accuracy $\e$ for evaluation of the rank, which makes sense in terms of physical implementation. Therefore the separation between ${\rm Sch}_{\e}(U_{od})$ and ${\rm ppr}_{\e}(U_{od})$ is at least quasipolynomial for a certain class of permutation matrix $U$, where the subscript $\e$ has the same meaning as above.

It is interesting to note that the problem of the separation of the rank and nonnegative rank is related to the log-rank conjecture  \cite{LovaszS88} in communication complexity theory (as remarked in \cite{Kol14}). It is curious that the nonlocal cost for implementing bipartite permutation unitaries (or reversible circuits) under our protocol is related to the communication complexity theory in this unexpected way.
\qed }
\eex

The lower bound mentioned above is not polynomial. So our protocol in the proof of Lemma~\ref{le:partial_permutation} is not efficient for some subclass of bipartite permutation unitaries represented by Eq.~\eqref{eq:perm_example}. There is an alternative method of implementing these unitaries, illustrated in Example~\ref{ex:example2} below, by noting that the local gate $\ketbra{2i-1}{2i} + \ketbra{2i}{2i-1}$ applied twice is the projector on the two-dimensional subspace spanned by $\ket{2i-1}$ and $\ket{2i}$. Rather than expanding $U_{od}$ using the form \eqref{eq:ppexpansion}, we can write $U$ as the product of some controlled-permutation unitaries controlled in the computational basis of $\cH_B$, where the controlled operators on $\cH_A$ are either $I_A$, or a permutation unitary which is the direct sum of an identity operator on a two-dimensional subspace, with another operator which is the tensor product of the identity operator on an $(M-k)$-dimensional space and the operator $\ketbra{1}{2}+\ketbra{2}{1}$ on a two-dimensional space. Each such controlled-permutation gate can be implemented using two bipartite CNOT gates and some local unitaries with the method in Lemma~\ref{le:bipartite_CNOT}. The nontrivial part of these controlled-permutation unitaries could overlap with each other. It corresponds to that in the expansion of $T$ in terms of binary vectors, the operator XOR is to be used instead of the usual ``+'' operator. That is, $T=\oplus_j (u_j \ox v_j)$, where $u_j$ and $v_j$ are binary column and row vectors, respectively, and $\oplus$ represents element-wise XOR operation (modulo-2 addition) of two or more matrices. Therefore, the relation between the XOR-rank of binary matrices and the rank of these matrices is relevant for the separation of the implementation cost and the Schmidt rank of $U$. In general, there is no definite inequality relation between the XOR-rank and the rank of a binary matrix \cite{xorrank}. Thus there may be some cases where this modified protocol is quite efficient, but in the other cases it is not too bad either, since $\rank_X(T)\le\rank_N(T)$ holds for any binary matrix $T$, meaning that it is better than the original protocol.

\bex\label{ex:example2}
{\rm
Let $U$ be of the form in Eq.~\eqref{eq:perm_example}, where $d_A=6$, $M=d_B=3$, and $C_1=\diag(1,1,0)$, $C_2=\diag(1,0,1)$, $C_3=\diag(0,1,1)$. Let $V=(X\oplus X\oplus I_2)\ox \diag(1,1,0) + I_6 \ox \diag(0,0,1)$, where $X=\ketbra{1}{2}+\ketbra{2}{1}$, and $I_n$ is the $n\times n$ identity matrix. Let $W=(I_2\oplus X\oplus X)\ox \diag(0,1,1) + I_6 \ox \diag(1,0,0)$. Then $U=VW$. Each of $V$ and $W$ can be implemented by two bipartite CNOT gates (ignoring local unitary gates; same as below). Hence $U$ can be implemented by four bipartite CNOT gates. In comparison, the protocol in the proof of Lemma~\ref{le:partial_permutation}, enhanced by doing nontrivial operations for the off-diagonal part $U_{od}$ only, requires $6$ bipartite CNOT gates (the partial permutation rank of $U_{od}$ is $q=3$, and a reduction by a factor of $3$ applies because the partial permutations are in-place).
The corresponding $T$ matrix is
\bea
T=\left(\begin{array}{ccc}
                     1 & 1 & 0 \\
                     1 & 0 & 1 \\
                     0 & 1 & 1
                   \end{array}\right).
\eea
It has rank $3$, and XOR rank $2$. This is the simplest example of a binary matrix that has its XOR rank less than the rank.
}
\eex

\section{The case with ancillas}
\label{sec:ancilla}

The use of ancillas of constant size has been seen in the previous section. In this section, we show that the use of ancillas of variable size (sometimes required to be initialized in fixed states) can be useful for reducing the controlled-gate cost $c(U)$ or the number of CNOT gates needed, but sometimes at the cost of modifying the $U$ (e.g., the tensor product $U\ox I_G$ instead of $U$ itself is used in Proposition~\ref{pp:cu_inequality} below).

\bpp\label{pp:CNOT_with_ancillas}
Any bipartite unitary $U$ on $d_A\times d_B$ system can be implemented by $4\lceil\log_2 \min\{d_A,d_B\}\rceil$ bipartite CNOT gates and some local unitaries, with the help of $\lceil\log_2 \min\{d_A,d_B\}\rceil$ ancilla qubits.
\epp
\bpf
The circuit for implementing $U$ is as follows: send the state of one system (which is embedded in an integer number of qubits) to the other party using $2\lceil\log_2 \min\{d_A,d_B\}\rceil$ CNOT gates, then perform the $U$ locally, and finally send the state of the said system back using the inverse of the first part of the circuit.  The first part of the circuit is a tensor product of many subcircuits each sending one qubit. Each such subcircuit is exactly the one-bit teleportation circuit in \cite{zlc00}, and requires one ancilla qubit which is initially in a fixed state and finally contains the one-qubit state being transferred. The number of subcircuits in the first part of the whole circuit is $\lceil\log_2 \min\{d_A,d_B\}\rceil$, and since final transfer back to the first system reuses the original qubits, no extra ancillas are needed, therefore the total number of ancilla qubits needed is $\lceil\log_2 \min\{d_A,d_B\}\rceil$.
\epf

Some special classes of bipartite unitaries can be implemented with small amounts of entanglement and classical communication \cite{ygc10}, which can also be expressed in terms of CNOT gates.  It should be noted that the upper bound $4\lceil\log_2 \min\{d_A,d_B\}\rceil$ is not optimal for all dimensions: as mentioned previously, in the case of $d_A=d_B=2$, only $3$ CNOT gates together with local unitaries are needed, without using ancillas. We do not know whether this upper bound is optimal for general unitaries in other dimensions, and this is listed as an open question in the next section.

Somewhat surprisingly, Example~\ref{ex:u_6gates} in Sec.~\ref{sec:Schmidt} shows the following:
\bpp\label{pp:cu_inequality}
For any bipartite unitary $U$, $c(U)\ge c(U\ox I_G)$, where $G$ is one qubit on the $A$ side. There are examples of $U$ satisfying $c(U)>c(U\ox I_G)$.
\epp
\bpf
The inequality is from observing that any decomposition of $U$ using controlled unitaries can be extended to a decomposition of $U\ox I_G$ with the same number of controlled unitaries. If $c(U)=c(U\ox I_G)$ always holds, we may repeatedly use it by adding one qubit on the $A$ side at a time, and get $c(U)=c(U\ox I_{A'})$, where $A'$ has an integer number of qubits and is of size at least as big as $A$. Then the method in Example~\ref{ex:u_6gates} implies that $c(U\ox I_{A'})\le 6$, thus $c(U)\le 6$, but this is generally impossible for a generic $U$ simply by parameter counting, see Sec.~\ref{sec:bipartite}. Therefore $c(U)=c(U\ox I_G)$ does not always hold.
\epf

\section{Conclusions}

\label{sec:con}

We have proposed the sandwich and generalized sandwich forms for the decomposition of bipartite and multipartite unitary operators, respectively. In particular, we have shown that any bipartite unitary on $\mathbb{C}^{d_A}\ox\mathbb{C}^{d_B}$ has a $(4d_A-5)$-sandwich form, and any $n$-partite unitary on $\mathbb{C}^{d_1}\ox\cdots\ox\mathbb{C}^{d_n}$ has a generalized $[2\prod^{n-1}_{j=1}(2d_j-2)-1]$-sandwich form. The numbers can be further reduced in some special cases. In particular, three controlled unitaries can implement a bipartite complex permutation operator. This last result can be applied to classical reversible circuits. We mentioned some connections between our results and the results in the literature. As an application of the types of decompositions above, we discussed how to express a bipartite unitary as the product of a simple type of bipartite gates and some local unitaries. We also discussed the relationship between the Schmidt rank of the unitary (bipartite permutation unitary in particular) and the complexity of the decomposition, and also discussed the use of local ancillas. To conclude this paper we present a few open questions by requiring that the gates are exactly implemented, and no ancillary space or system is allowed unless stated otherwise.
\bem
\item
Let $s(U)$ be the smallest number of controlled unitary gates required in a decomposition of $U$ of the sandwich form. Then $s(U)\ge c(U)$. Do we have $s(U)=c(U)$? We suspect that this does not hold for some $U$. But does the similar equality $\max_{U\in \cT_{ab}} s(U)=\max_{U\in \cT_{ab}} c(U)$ hold, where $\cT_{ab}$ is the set of all bipartite unitaries $U$ on an $a\times b$ dimensional space?
\item
Can we obtain some form of decomposition of bipartite unitaries in terms of controlled unitaries, by taking a hint from the decomposition of single-party unitary matrices in \cite[Corollary 1]{iw14} ? In particular, can we replace $4d_A-5$ by $2d_A-1$ in \eqref{eq:upperbound}?
\item
Let $U$ be a bipartite unitary on the $d_A\times d_B$ system. Do the following equations hold?
\bea\label{eq:questions1}
&c(U)=c(U+\proj{d_A+1}\ox I_B),\notag\\
&c_s(U)=c_s(U+\proj{d_A+1}\ox I_B),\\
&c_e(U)=c_e(U+\proj{d_A+1}\ox I_B).\notag
\eea
\item
It is obvious that the following inequalities hold:
\bea\label{eq:questions2}
c(U^{\ox n})&\le& c(U),\notag\\
c_s(U^{\ox n})&\le& n\cdot c_s(U),\\
c_e(U^{\ox n})&\le& n\cdot c_e(U).\notag
\eea
Here the bipartite unitary $U^{\ox n}=U_{A_1B_1}\ox\cdots\ox U_{A_nB_n}$ acts on the space $\cH_A\ox\cH_B$ where $\cH_A=\bigox^n_{i=1}\cH_{A_i}$ and $\cH_B=\bigox^n_{i=1}\cH_{B_i}$. But do the equalities always hold in the three inequalities above?
\item
As discussed in Sec.~\ref{sec:elementary}, a bipartite permutation unitary can be implemented using a certain number of standard gates of the CNOT type and some local transposition gates. What is the minimum number of these gates needed to implement any bipartite permutation unitary on $d_A\times d_B$ space?  And since the local gates can be regarded as easy to implement, we can also ask the following: What is the minimum number of the first type of gates needed?
\item
For $d_A$ or $d_B$ greater than 2, can an upper bound better than $4\lceil\log_2 \min\{d_A,d_B\}\rceil$ be found for the number of CNOT gates (or the bipartite elementary gates defined in Definition~\ref{def:ele2}) needed to decompose a bipartite unitary with the help of local ancillas and local unitaries?
\smallskip
\item
Is there a dimension-independent upper bound of $c(U)$ in terms of the Schmidt rank $r$ of a general bipartite unitary $U$? It is known that $c(U)=1$ when $r=2$ or $3$ \cite{cy13,cy14ap}. See also the discussions in Sec.~\ref{sec:Schmidt}, and Theorem~\ref{thm:u_permutation} (which is for the permutation unitaries only and requires ancillas of fixed size).
\smallskip
\item
For given integers $m$, $n$ satisfying $m>n\ge 2$, is there a dimension-independent upper bound (as a function of $m,n$ only) of the number of Schmidt-rank-$n$ bipartite unitaries needed to decompose any Schmidt-rank-$m$ bipartite unitary on the same space?
What about restricting the target unitary to be a controlled unitary in this question?

\eem

\smallskip
\section*{Acknowledgments}

We thank Joseph Fitzsimons and Kae Nemoto for helpful discussions.  We also thank Scott Cohen for careful reading of an early version of the paper and pointing out a couple of issues in the presentation. This material is based on research funded in part by the Singapore National Research Foundation under NRF Grant No. NRF-NRFF2013-01, and in part by NICT-A (Japan).

\bibliographystyle{unsrt}

\bibliography{channelcontrol}

\begin{thebibliography}{10}
\bibitem{note1}
Throughout the paper a local unitary is in general not a single-qubit unitary;
  ``local'' is with respect to the system partition.
\end{thebibliography}

\end{document}